\documentclass[%
 aip,
 amsmath,amssymb,
 reprint,%
]{revtex4-1}

\usepackage{graphicx}
\usepackage{dcolumn}
\usepackage{bm}

\usepackage[utf8]{inputenc}
\usepackage[T1]{fontenc}
\usepackage{mathptmx}
\usepackage{etoolbox}
\usepackage{mhchem}

\makeatletter
\def\@email#1#2{%
 \endgroup
 \patchcmd{\titleblock@produce}
  {\frontmatter@RRAPformat}
  {\frontmatter@RRAPformat{\produce@RRAP{*#1\href{mailto:#2}{#2}}}\frontmatter@RRAPformat}
  {}{}
}%
\makeatother

\usepackage{color}

\begin{document}

\preprint{AIP/123-QED}

\title{A static quantum embedding scheme based on coupled cluster theory}
\author{Avijit Shee}
\affiliation{Department of Chemistry, University of California, Berkeley, CA  94720, USA}
\email{ashee@berkeley.edu}
\author{Fabian M. Faulstich}
\affiliation{Department of Mathematics, Rensselaer Polytechnic Institute, Troy, NY 12180, USA}
\author{Birgitta Whaley}
\affiliation{Department of Chemistry, University of California, Berkeley, CA  94720, USA}
\affiliation{Berkeley Center for Quantum Information and Computation, Berkeley, CA 94720, USA}
\author{Lin Lin}
\affiliation{Department of Mathematics, University of California, Berkeley, CA 94720, USA}
\affiliation{Applied Mathematics and Computational Research Division, Lawrence Berkeley National Laboratory, Berkeley, CA 94720, USA}
\author{Martin Head-Gordon}
\affiliation{Department of Chemistry, University of California, Berkeley, CA 94720, USA}
\affiliation{Chemical Sciences Division, Lawrence Berkeley National Laboratory, Berkeley, CA 94720, USA}

\date{\today}

\begin{abstract}
We develop a static quantum embedding scheme that utilizes different levels of approximations to coupled cluster (CC) theory for an active fragment region and its environment. To reduce the computational cost, we solve the local fragment problem using a high-level CC method and address the environment problem with a lower-level M{\o}ller-Plesset (MP) perturbative method. This embedding approach inherits many conceptual developments from the hybrid MP2 and CC works by Nooijen and Sherrill (J. Chem. Phys. 111, 10815 (1999), J. Chem. Phys. 122, 234110 (2005)). We go beyond those works here by primarily targeting a specific localized fragment of a molecule and also introducing an alternative mechanism to relax the environment within this framework. We will call this approach MP-CC. We demonstrate the effectiveness of MP-CC on several potential energy curves, and a set of thermochemical reaction energies, using CC with singles and doubles as the fragment solver, and MP2-like treatments of the environment. The results are substantially improved by the inclusion of orbital relaxation in the environment. Using localized bonds as the active fragment, we also report results for \ce{N=N} bond breaking in azomethane and for the central \ce{C-C} bond torsion in butadiene. We find that when the fragment Hilbert space size remains fixed (e.g., when determined by an intrinsic atomic orbital approach), the method achieves comparable accuracy with both a small and a large basis set. Additionally, our results indicate that increasing the fragment Hilbert space size systematically enhances the accuracy of observables, approaching the precision of the full CC solver. 
\end{abstract}

\maketitle

\section{Introduction}

Coupled cluster (CC) theory\cite{Cizek_CC} provides the best balance between efficiency and accuracy for calculating the electronic structure of weak to moderately correlated molecules. However, CC theory with even the simplest singles and doubles truncation, still has computational cost scaling as $\mathcal{O}(O^2 V^4)$, where $O$ and $V$ are the numbers of occupied and virtual orbitals being correlated. This precludes calculations on large molecular systems using large basis sets. At a lower level of accuracy, such problems are often addressed with composite methods. There is a long history of such methods in quantum chemistry, including the Gaussian-$n$ methods (G1-G4)\cite{G1JCP1989, G2JCP1991, G3JCP1998, G4JCP2007}, the Weizmann-$n$ methods (W1-W4)\cite{W12JCP1999, W3JCP2004, W4JCP2006}, the HEAT protocol\cite{HEATJCP2004}, the correlation consistent Composite Approach (ccCA)\cite{WilsonJCP06, WilsonMolPhys09}, etc. As a simple example, the basis set correction to an expensive coupled cluster CC calculation such as CCSD(T) might be approximated using a far less demanding approach such as second-order M{\o}ller-Plesset theory (MP2) that exhibits similar basis set dependence via:
\begin{equation}
    E_{\mathrm{Composite}} \approx E_{\mathrm{CC/small}} + \left( E_{\mathrm{MP2/large}} - E_{\mathrm{MP2/small}} \right) .\label{Eq:composite}
\end{equation}
In a similar additive fashion, the ONIOM \cite{ONIOM_JCC95, ONIOM_review_ChemRev15} model treats different parts of a system with different levels of theory to tackle the system size scaling.  

Local correlation methods \cite{PulayReview93, Subotnik_lcc06, PNO_CC_Neese13, Werner_review_2018} are a more advanced treatment of the system-size scaling that exploits the locality of electron correlations to approach linear scaling of compute effort with size.
In contrast, our goal is to approximate the total energy using a high-level treatment for a specific local region of a big molecule and to control the computational expenses of such calculations by employing a lower-level approximation for the environment. This ``QM-in-QM'' approach is useful for addressing the chemistry of an active region of a catalyst, or point defects in materials, for example. Or, when the active fragment spans all the atoms in the molecule, a similar approach can be used to address the basis set convergence of the total problem. This will serve as an alternative method to overcome the bottleneck of large basis set calculations with CC, for which many methods have been already proposed, such as explicitly correlated R12/F12 methods \cite{CCSDT-F12_kniziaJCP09}, diagrammatic separation-based extrapolation \cite{ppl_extrapolation_JCP19}, and cardinal number based extrapolation \cite{Helgaker_JCP97}, among many others.     

Our effort to address the challenges of large system size and large basis set calculations can also be viewed from a quantum embedding perspective. In quantum embedding, one identifies an important (preferably local) region of the system, which is denoted as the fragment (F). The remainder of the system is referred to as the environment (E) and is treated by a low-scaling method, such as Hartree-Fock, density functional theory, or a perturbative method. For the fragment problem, one first evaluates a coupling potential ($\Delta$) between the environment and the fragment. This coupling potential is then incorporated into the fragment Hamiltonian treating the fragment problem as an open-quantum system. The fragment problem is solved with a high-accuracy method (typically called a solver). CC methods have proven to be successful as a solver for many embedding methods, especially when the fragment size needs to be large \cite{dmet_cuprate_science, Shee2019, Zhu_DMFT_PRX21, Yeh_SheePRB21}. 
This highly accurate solution on the fragment is subsequently used to self-consistently update $\Delta$. 
The fragment problem will subsequently be described by a renormalized interaction, and solved by a CC method truncated at a certain excitation rank. The environment problem 
is then treated by a lower-scaling method such as MP2 theory \cite{werner_LMP2_JCP15, Neese_Valeev_LMP2_JCP15, Zhenling_MHG_JCP23}. The cost of the fragment problem remains constant with the increase of system size/basis set size and thus reduces the overall cost. 

Here we combine the fragment and the environment problems using two sets of coupled projective equations: one for the environment and one for the fragment. Furthermore, we invoke a self-consistency condition between the fragment and the environment based on the ``global'' CC amplitudes. This approach inherits many essential ideas from the hybrid MP2 and CCSD methods developed by Nooijen \cite{nooijen1999combining} and Sherrill et al.~\cite{bochevarov2005hybrid,bochevarov2006hybrid} Both Nooijen and Sherill \textit{et al.} presented two versions of the theory. Their simplest version did not relax orbitals in the environment; however, in an improved version such contributions were included. In that version, certain classes of environment amplitudes (external and semi-internal) were updated using the same amplitude equations as CCSD. These amplitudes were chosen such that the computational cost remains low. This idea has its origin in earlier work by Adamowicz, Piecuch, and coworkers published between 1993 and 1998, which introduced partitioning of the cluster operator into internal and external components.~\cite{piecuch1993state,piecuch1994state} 

Our approach differs from Nooijen and Sherrill \textit{et al.}'s approach in two crucial aspects. First, instead of choosing canonical molecular orbital (CMO)s as a basis for the active space, we choose localized MO (LMO)s, which lets us define the active fragment as a bond or region. Second, the relaxation of the environment orbitals is performed differently (vide infra) with the objective in mind that it does not increase the cost of the low-level method even when we increase the excitation rank of the high-level CC method. Another related approach is the multi-level CC (MLCC) theory of Koch and co-workers~\cite{MLCCJCP2014, MLCCjctc2021}. This approach also utilizes a localized active space and also defines a mechanism for relaxing the environment. Differences between our approach and the MLCC models in terms of the general framework and computational implications will be elaborated in Sec.~\ref{sec:method_comparison} and Appendix~\ref{Sec:complexity_triples}. 


The remainder of the manuscript is organized as follows.
In Section~\ref{sec: theory}, we describe the general framework, then outline various choices for the environmental projection equation, and finally discuss some theoretical aspects of the fragment amplitude equations. We will use CCSD as the fragment solver, although this choice can be extended to higher-level CC models.
As in all embedding methods, it is important to choose a suitable local basis that describes the fragment problem well, as the results 
are not invariant with respect to this choice. Our choice of localized representation is described in Section~\ref{sec:active_space_basis}. To follow up on the review of existing embedding methods given above, in Section~\ref{sec:method_comparison} we compare some formal properties of our current method with related methods. 
Implementation details are briefly summarized in Section~\ref{sec:Implimentation}.
Finally, in Section~\ref{sec:results} we will illustrate various aspects of the theory via a set of prototypical numerical examples. We will compare different versions of our embedding theory relative to the full high-level method, and a composite approach, which can be regarded as an uncoupled version of the embedding method.  

\section{Embedded coupled cluster formalism}
\label{sec: theory}

Coupled cluster theory is a wave function approach that expresses the ground-state wave function using an exponential parametrization\cite{hubbard1957description,hugenholtz1957perturbation}. Given a single Slater determinant reference state like the Hartree-Fock (HF) determinant $|\Phi_0\rangle$, any intermediately normalized state $|\Psi\rangle$ obeying $\langle \Phi_0 | \Psi \rangle = 1$ can be uniquely written as $e^T|\Phi_0\rangle$, where $T$ is the corresponding cluster operator. To define the cluster operators, we introduce the notation that $i$, $j$, $k$, $l$... label the occupied orbitals and $a$, $b$, $c$, $d$,... the virtual orbitals. The cluster operator is then given by
\begin{equation}
T 
= \sum_{\mu \in \mathcal{I}} t_\mu X_\mu
= \sum_{k=1}^N \sum_{\substack{\mu \in \mathcal{I}\\ |\mu| = k}} t_\mu X_\mu,
\end{equation}
where 
\begin{equation}
\mathcal{I}
=
\left\lbrace
{a_1,...,a_k \choose i_1,...,i_k}
~:~1\leq k\leq N
\right\rbrace
\end{equation}
and $X_\mu$ are particle-hole excitation operators \cite{Cizek_CC}.
The CC amplitudes, denoted by ${\bf t}= (t_\mu)_{\mu \in \mathcal{I}}$, are determined through projective equations:

\begin{equation}
\label{Eq: CC_ampl}
\langle \Phi_\mu| \overline{H} | \Phi_0 \rangle =  0 \qquad \forall \mu \in \mathcal{I},
\end{equation}
where 
\begin{equation}
\label{eq:SimHam}
\overline{H}=e^{-T} H e^{T}
\end{equation}
and $\langle \Phi_\mu| = \langle \Phi_0| X_\mu^\dag$ denotes an excited Slater determinant with an excitation rank $|\mu|>0$. The corresponding CC energy is then given by 
\begin{equation}
\label{Eq: CC_energy}
E_{\rm CC} = \langle \Phi_0 | \overline{H} | \Phi_0 \rangle.
\end{equation}


Since the untruncated CC equations~\eqref{Eq: CC_ampl} rapidly become numerically intractable, truncations are commonly employed. The subject of this work is the CCSD variant,
where $T$ comprises excitation operators up to a maximum excitation rank of two, i.e., 
\begin{equation}
T = 
\sum_{i,a} t_i^a X_i^a
+ \frac{1}{4} \sum_{i,j,a,b}  t_{ij}^{ab} X_{ij}^{ab}. 
\end{equation}

However, even CCSD has computational effort scaling as $\mathcal{O}(O^2 V^4)$, where $O$ and $V$ denote the number of occupied and virtual orbitals included in the correlation treatment. Even CCSD becomes computationally demanding when naively applied to larger systems.

Here we explore an approximation to CCSD that is inspired by quantum embedding theory. The underlying core concept is well-known: it is the recognition that certain CC amplitudes possess greater physical significance than others if expressed in a suitable basis, such as localized or natural orbitals.
The proposed approach aims to identify and accurately evaluate these critical amplitudes, while the less significant amplitudes are approximated via many-body perturbation theory with reduced precision (vide infra). 
We therefore name this approach the many-body perturbation coupled-cluster (MP-CC) method. It aims to maintain high accuracy while simultaneously reducing overall computational complexity. 

Underlying the MP-CC approach is a partition of the total orbital space into two sets: the active fragment (F) orbitals and the environment (E) orbitals. The fragment orbitals describe the chemically relevant region that requires higher accuracy. The fragment orbitals are formed from intrinsic atomic orbitals (IAOs) \cite{IAO_knizia_JCTC13} which span only a small chemically relevant subspace of the full basis set of the calculation. The environment orbitals, on the other hand, will describe additional virtual orbitals associated with the full atomic orbital basis, and all occupied orbitals not contained in the fragment orbital space. This allows the use of all valence occupied and some virtual orbitals as the fragment. Alternatively, the fragment can be just a single local site of particular interest (e.g.~a catalytic active site), or a set of local sites.

A valid partition splits the occupied and virtual orbitals into environment and fragment subsets, such that
\begin{equation}
O = O_{\rm E} + O_{\rm F} 
\quad {\rm and} \quad 
V = V_{\rm E} + V_{\rm F}.
\end{equation}
Once a partition is made, there are also two classes of cluster operators: $T_{\rm F}$, and $T_{\rm E}$. Note that the $T_{\rm E}$ operators contain amplitudes that involve only environment orbitals, as well as mixed amplitudes that contain both fragment and environment orbitals. The latter describes correlations between electrons in fragment and environment occupied levels, or between two occupied fragment levels using environment virtuals, etc. 

In the spirit of quantum embedding, we will compute the cluster operators $T_{\rm F}$, and $T_{\rm E}$ using different levels of sophistication. More precisely, we propose to determine the fragment cluster operator $T^{\rm F}$ using the standard CC formalism (the high-level theory), involving a full expansion of $\overline{H}$ (see Eq.~\eqref{eq:SimHam}) to yield $\overline{H}_{\rm F}^{\rm high} = \overline{H}$.
In contrast, the calculation of $T_{\rm E}$ is based on a lower-order perturbative expansion of $\overline{H}$, resulting in $\overline{H}_{\rm E}^{\rm low}$ (vide infra). This lower-order expansion offers various options for gaining a computational advantage, as will be elaborated in Sec.~\ref{sec:h_low}. The MP-CC approach results in the following set of coupled projective equations: 
%
%
\begin{align}
     \langle \Phi_\mu^{\rm F}| \overline{H}_{\rm F}^{\rm high}| \Phi_0 \rangle = {}& 0, \label{Eq:projF} \\
     \langle \Phi_\mu^{\rm E}| \overline{H}_{\rm E}^{\rm low} | \Phi_0 \rangle = {}& 0. \label{Eq:projEB}
\end{align}
We emphasize that both $\overline{H}_{\rm F}^{\rm high}$ and $\overline{H}_{\rm E}^{\rm low}$ in Eqs.~\eqref{Eq:projF} and~\eqref{Eq:projEB} contain the full set of $T$-amplitudes i.e., $\mathbf{t}_\mathrm{F}$ and $\mathbf{t}_\mathrm{E}$. In other words, although Eq.~\eqref{Eq:projF} solves only for ${\bf t}_{\rm F}$, the ${\bf t}_{\rm E}$ amplitudes still enter the equations. Similarly, Eq.~\eqref{Eq:projEB} solves for ${\bf t}_{\rm E}$ in the presence of ${\bf t}_{\rm F}$. 
Therefore, Eqs.~\eqref{Eq:projF} and \eqref{Eq:projEB} should be solved self-consistently, unless they are uncoupled by the perturbative approximations.
The total CC energy remains as Eq.~\eqref{Eq: CC_energy}.

Moreover, the energy and amplitude equations of the MP-CC approach can be combined into a Lagrangian:
\begin{equation}
\begin{aligned}
\mathcal{L}({\bf t},\boldsymbol{\lambda}) &=
\langle \Phi_0| \overline{H} | \Phi_0 \rangle 
+ \langle \Phi_0 | \Lambda_{\rm E} \overline{H}_{\rm E}^{\rm low} | \Phi_0 \rangle\\
& + \langle \Phi_0 | \Lambda_{\rm F} \overline{H}_{\rm F}^{\rm high} | \Phi_0 \rangle, \label{Eq:lagrangian}
\end{aligned}    
\end{equation}
where $ \Lambda = \sum_{\mu\in \mathcal{I}} \lambda_\mu X_\mu^{\dag}$ consists of the dual variables $\lambda_\mu$, ensuring that Eqs.~\eqref{Eq:projF} and \eqref{Eq:projEB} are satisfied; $\Lambda_{\rm E}$ and $\Lambda_{\rm F}$ are defined in accordance with  $T_{\rm E}$ and $T_{\rm F}$. 
The working equations then correspond to a first-order optimality condition of the Lagrangian, i.e., 
\begin{equation}
\label{eq:Lagrangian}
\left\lbrace
\begin{aligned}
\frac {\partial\mathcal{L}}{\partial {\bf t}_\alpha} ={}& 0, \qquad \alpha\in\left\{\rm E,F\right\}\\
\frac {\partial\mathcal{L}}{\partial \boldsymbol{\lambda}_\alpha}={}& 0, \qquad \alpha\in\left\{\rm E,F\right\}.
\end{aligned}
\right.
\end{equation}
Compared to other embedding approaches, the Lagrangian in Eq.~\eqref{Eq:lagrangian} enables us to access a broad range of observables beyond the energy, by employing analytic energy derivative techniques~\cite{CIenergyderJCP1984,CCenergyderJCP1990}.


\subsection{Environment amplitude equations}
\label{sec:h_low}

We employ two distinct approaches to the lower-order expansion $\overline{H}_{\rm E}^{\rm low}$, both based on perturbative arguments. 
\subsubsection{Unrelaxed approach}\label{sec:h_low_un}

In the first approach, we use the simplest possible MP2-like projection equations for the environment amplitudes, i.e.,
\begin{equation}
\langle \Phi_\mu^{\rm E}| V + [F,T] | \Phi_0 \rangle = 0. \label{Eq:MP2}
\end{equation}
While $T$ contains all cluster amplitudes, Eqs.~\eqref{Eq:MP2} are used to determine only $T_{\rm E}$. In defining these equations, we have used the partitioned Hamiltonian: $H = F + V$, where $F$ is the mean-field Hamiltonian or Fock operator from HF theory and $V$ is the two-particle fluctuation potential, expressed in the HF basis. Note that we use normal-ordered operators for $F$ and $V$, though we do not denote it explicitly. By the Brillouin theorem, $\langle \Phi_s^{\rm E}| V | \Phi_0 \rangle = 0$ for all singles, $|\Phi_s^{\rm E} \rangle$ so the environment orbitals are unrelaxed by singles amplitudes.  We therefore refer to Eqs.~\ref{Eq:MP2} as the unrelaxed MP-CCSD method. Denoting fragment singles as $S_\mathrm{F}$, fragment doubles as $T_\mathrm{F}$ and environment doubles as $T^{(1)}_\mathrm{E}$, this is a wavefunction ansatz of the form:
\begin{equation}
   |\Psi_\mathrm{MP-CC(unrelaxed)} \rangle = e^{S_\mathrm{F}} e^{T^{(1)}_\mathrm{E}+T_\mathrm{F}} | \Phi_0\rangle. 
\end{equation}
The unrelaxed method yields fixed environment amplitudes that influence the fragment amplitudes but not vice-versa.

\subsubsection{Relaxed approach} \label{sec:h_low_re}
In the second approach, we relax the wave function in the sense of Thouless relaxation~\cite{thouless1960stability}, wherein the HF determinant $\Phi_0$ can be transformed into another determinant $\Phi'_0$ by the exponential of a single substitution operator, denoted as $S$: $\Phi'_0 \propto e^S \Phi_0$. 
Higher-body amplitudes describing correlation, such as doubles will still be considered only to first order. We have made this distinction apparent by denoting singles amplitudes as $S \equiv S_\mathrm{F} + S_\mathrm{E}$, and environment doubles amplitudes as $T^{(1)}_\mathrm{E}$ in the relaxed MP-CC wave function ansatz 

\begin{equation}
|\Psi \rangle = e^S e^{T^{(1)}_\mathrm{E}+T_\mathrm{F}} | \Phi_0\rangle.
\end{equation}

More precisely, we will assign the formal perturbation parameter $\eta$ to the interaction term of the similarity transformed Hamiltonian: $\Tilde{H} = e^{-S} H e^S$ , that is, $\Tilde{H} = E_{cl} + \Tilde{F}^{(0)} + \eta \Tilde{V}^{(1)}$. Here, $\Tilde{F}$ and  $\Tilde{V}$ are the one-particle and two-particle parts of the transformed Hamiltonian, respectively, and $E_{cl}$ is a closed scalar contribution, which will not contribute to the projective equations. 
Keeping the terms up to first order in the perturbation, $\eta$, yields the following projective equations for the singles (s) and doubles (d) amplitudes:

\begin{align}
    \langle \Phi_s^{\rm E}|  \Tilde{F} + [ \Tilde{F},T^{(1)}_{\rm E}] + [ \Tilde{F},T_{\rm F}]  | \Phi_0 \rangle ={}& 0, \label{Eq:low_relax_singles}  \\
    \langle \Phi_d^{\rm E}|  \Tilde{V} + [ \Tilde{F},T^{(1)}_{\rm E}] + [ \Tilde{F},T_{\rm F}]  | \Phi_0 \rangle ={}& 0.  \label{Eq:low_relax_doubles}
\end{align}
\noindent In this work, $T^{(1)}_{\rm E}$ contains the first-order environment doubles amplitudes, while the fragment doubles amplitudes are in $T_{\rm F}$. 
In the derivation of Eq.~\eqref{Eq:low_relax_singles} and~\eqref{Eq:low_relax_doubles} we used the fact that the similarity transformation with $e^S$ does not increase the operator rank of the Hamiltonian.
\noindent By comparing Eqs.~\eqref{Eq:low_relax_singles} and~\eqref{Eq:low_relax_doubles} with Eq.~\eqref{Eq:projEB}, we obtain the definition of $\overline{H}_{\rm E}^\mathrm{low}$. 

In Fig.~\ref{fig:embed_scheme}, we schematically show the workflow of the proposed method ($W_{\rm F}$ is defined in \ref{sec:Fragampl}, and not relevant for the following discussion). 
Note that when employing the relaxed method (i.e., the second approach outlined above) in step 3 (see Fig.~\ref{fig:embed_scheme}), the first iteration uses $S=0$ implying that Eq.~\eqref{Eq:low_relax_singles} and~\eqref{Eq:low_relax_doubles} reduce to Eq.~\eqref{Eq:MP2}. In particular, in the first iteration, the two procedures outlined above yield the same inactive amplitudes. However, in the subsequent iterations, non-zero active singles amplitudes generated from the $T_{\rm F}$-amplitude equations will contribute to Eq.~\eqref{Eq:low_relax_singles} and~\eqref{Eq:low_relax_doubles}, thus yielding non-zero inactive singles amplitudes.

The computational scaling of Eq.~\eqref{Eq:low_relax_singles} and \eqref{Eq:low_relax_doubles} is MP2-like, using a factorized two-body Hamiltonian, as only low-rank singles amplitudes contract with them. The factorization of the two-body Hamiltonian can be carried out using either the density fitting (DF) \cite{Whitten_DF_JCP73, Dunlop_DF_JCP79} approach or a cholesky decomposition (CD) \cite{Koch_CD_JCP2000} approach. Following Mester \textit{et. al.} \cite{MesterCC2_JCP17}, this yields $\mathcal{O}(O^2 V^2 X)$ computational scaling for our relaxed approach, where $X$ is the dimension of the auxiliary basis used in the DF or CD scheme.

\subsection{Fragment amplitude equations} \label{sec:Fragampl}

An important observation is that in the fragment amplitude equations~\eqref{Eq:projF}, the Hamiltonian $\overline{H}_{\rm F}^\mathrm{high}$ is screened by the environment amplitudes ${\bf t}^{\rm E}$.
We may rewrite Eq.~\eqref{Eq:projF} in the following way: 
\begin{equation}
    \langle \Phi_\mu^{\rm F}| e^{-T_{\rm F}}W_{\rm F} e^{T_{\rm F}} | \Phi_0 \rangle = 0,\label{Eq:projFmod}
\end{equation}
where
\begin{equation}
\label{Eq:WF}
\begin{aligned}
   W_{\rm F} ={}& [e^{-T_{\rm E}} H e^{T_{\rm E}} ]_ {\rm F} \\
        ={}& W_{1b, {\rm F}} + W_{2b,  {\rm F}} + W_{3b,  {\rm F}} + ...    
\end{aligned}
\end{equation}
In Appendix \ref{Sec:WFEquations} we show the working equations arising from Eq.~\eqref{Eq:projFmod}.

\begin{figure}
    \centering
    \includegraphics[width=0.47\textwidth]{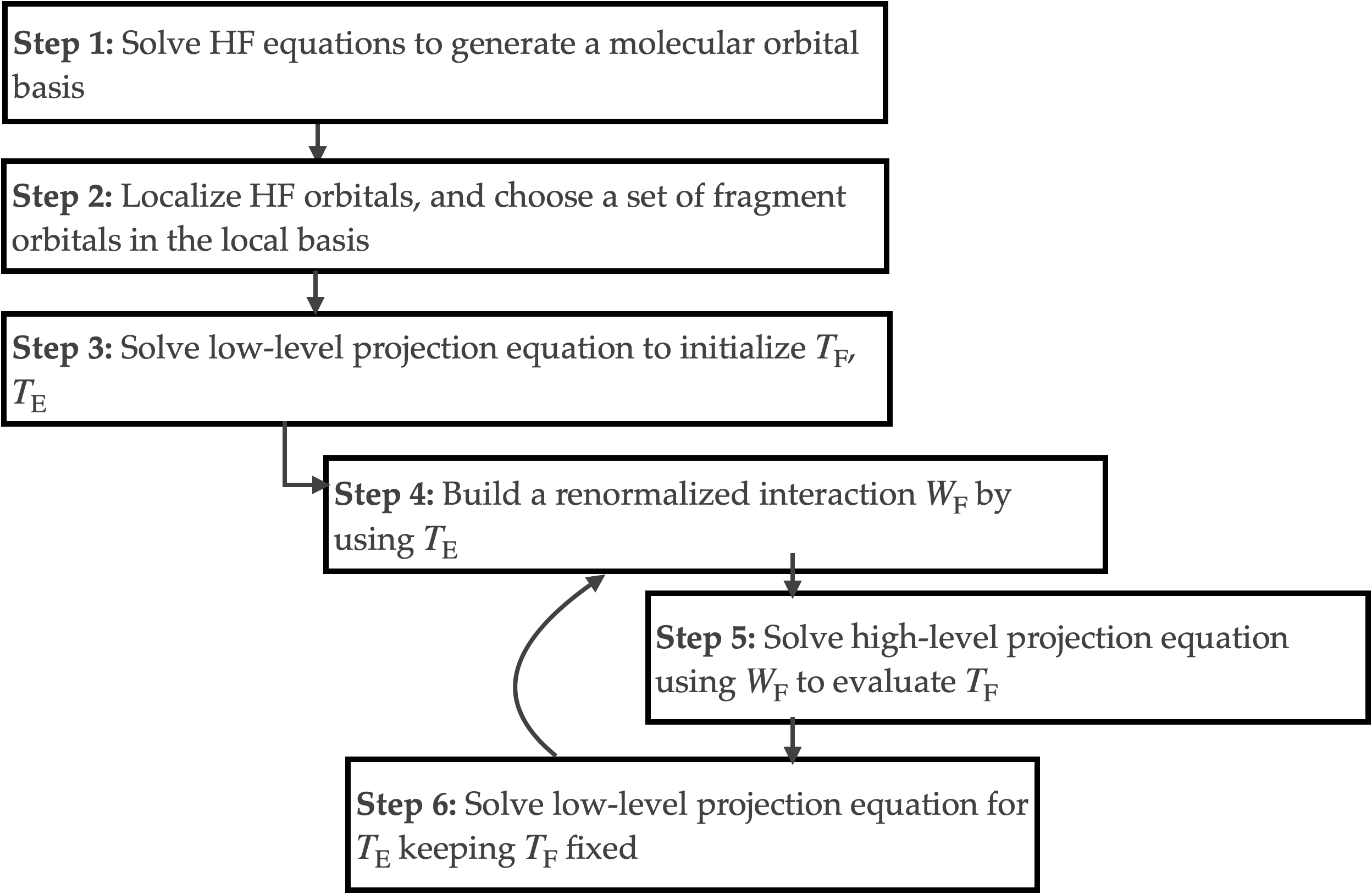}
    \caption{Schematic view of the current embedding scheme.}
    \label{fig:embed_scheme}
\end{figure}

In Eq.~\eqref{Eq:WF}, we renormalize the bare Hamiltonian by applying a similarity transformation via $e^{T_{\rm E}}$ and then restrict the orbital indices to the fragment.
Note that the similarity transformation increases the rank of $W_{\rm F}$ compared to the bare Hamiltonian. However, when solving Eq.~\eqref{Eq:projFmod}, we order the operation of the tensors such that we avoid generating higher rank tensors of $W_{\rm F}$. 
This particular construction is conceptually important because it is the mechanism by which the low-energy fragment (or active region) Hamiltonian is screened by the ``high-energy'' environment amplitudes, which is a crucial requirement for accurate quantum embedding theories \cite{vanLoonPRB2021,WernerPRB2021}. We note that in a similar manner, Kowalski~\textit{et.~al.} constructed a CC downfolded Hamiltonian by using sub-system embedding sub-algebras (SES)~\cite{cc_downfold_kowalski23}.
In Fig.~\ref{fig:embed_scheme}, we schematically show that in step~4 and step~5, we first construct $W_{\rm F}$, and then iteratively solve Eq.~\eqref{Eq:projFmod}. The maximum complexity of solving Eq.~\eqref{Eq:projFmod} for the fragment amplitudes is $\mathcal{O}$($O_{\rm F}^2 V_{\rm F}^4$). If there are multiple (disjoint) fragments, it also scales linearly with the number of fragments. The construction of $W_{\rm F}$, on the other hand, is more computationally demanding, and scales as $\mathcal{O}$($O^2 O_{\rm F} V^2 V_{\rm F}$); however, this cost is non-iterative.


\section{Choice of basis set for active space} \label{sec:active_space_basis} 
 The choice of a suitable basis to define the active space is crucial for embedding methods. The dimensionality and accuracy of the fragment impurity problem are dependent on this choice, because the basis determines spatial locality on the one hand, and the most suitable virtual orbitals to capture the main electron correlation effects in the fragment region on the other hand. To capture spatial locality, localized orbitals are of course essential. To capture the most relevant virtuals for electron correlation, without performing an active space calculation, we adopt the Atomic Valence Active Spaces (AVAS)~\cite{AVASknizia2017} approach. AVAS straightforwardly defines valence occupied and virtual orbitals for a fragment, and is trivially able to incorporate higher angular momentum functions into the virtual space as discussed below. AVAS has important advantages over simply using canonical MOs, as illustrated in Appendix ~\ref{sec:cmovslmo}.
 
 The objective of AVAS is to isolate a set of valence/semi-valence atomic orbitals ($A$) denoted by $p,q$, which has overlap with the HF occupied ($|i\rangle$) and virtual orbitals ($|a\rangle$) of the full molecule. \\
 
 To this end, we define the projection operator 
\begin{equation}
    P_A = \sum_{p, q \in A} [\sigma^{-1}]_{pq} |p\rangle \langle q | \quad {\rm where}\quad \sigma_{pq} = \langle p | q \rangle.
\end{equation}
With this projection operator, we then construct two overlap matrices 
\begin{equation}
[S^A]_{ij} = \langle i | P_A|j \rangle 
\quad {\rm and}\quad 
[S^A]_{ab} = \langle a | P_A|b \rangle
\end{equation}
in the space of occupied and virtual orbitals, respectively. To isolate the active orbitals, $[S^A]_{ij}$ and $[S^A]_{ab}$ are diagonalized and the orbitals ${|i\rangle}$ and ${|a\rangle}$ are rotated by the diagonalizing matrices to generate two new sets of orbitals: ${|\Tilde{i}\rangle}$ and ${|\Tilde{a}\rangle}$.
We then choose the active orbitals to be the vectors in ${|\Tilde{i}\rangle}$ and ${|\Tilde{a}\rangle}$ corresponding to non-zero eigenvalues. In the construction of $[S^A]_{ij}$ matrix, we freeze the core occupied orbitals, which will then be added to the inactive (environment) orbitals. 


The atomic orbitals defining the fragment must be specified before the above procedure can be invoked. Commonly, a minimal atomic orbital basis (e.g., such as MINAO\cite{IAO_knizia_JCTC13} or autogenerated via AutoSAD\cite{gimferrer2022oxidation}) is employed for constructing $A$. 
An important characteristic of the AVAS scheme is that one can alternatively employ a slightly larger reference basis set so that the active space includes orbitals of higher angular momentum. This aspect is important, for instance, in first-row transition metals, where often ``double-shell'' or 4d orbitals are required.
We will project the AVAS reference basis such that it exactly spans an integer number of fragment occupied and virtual orbitals equal to its dimension in our system. This restriction is especially crucial for the occupied space. 

\section{Comparison with existing approaches}\label{sec:method_comparison}
In the hybrid MP2 and CCSD approaches developed by Nooijen \cite{nooijen1999combining} and Sherrill et al.\cite{bochevarov2005hybrid}, two very similar unrelaxed schemes were introduced, namely, A-CC/PT and MP2-CCSD(I). Our unrelaxed scheme, as defined in Sec.~\ref{sec:h_low_un}, is identical to those approaches for a given choice of fragment orbitals. However, in contrast to the canonical molecular orbitals (CMOs) chosen in their implementations, we choose localized molecular orbitals (LMOs) from the AVAS scheme (Sec.~\ref{sec:active_space_basis}) to construct the active space, which has the advantage of treating a spatially localized fragment. Furthermore, both Nooijen and Sherrill \textit{et al.} defined two improved versions of their methods, namely R-CC/PT and MP2-CCSD(II), respectively by including selected sets of external and semi-internal singles and semi-internal doubles amplitudes. In our approach, on the other hand, we relax the whole set of environment amplitudes (which corresponds to external plus semi-internal amplitudes) with a low-level method, leading to Eq.~\eqref{Eq:low_relax_singles} and \eqref{Eq:low_relax_doubles}.

In the Multi-Level CC (MLCC) work by Koch and co-workers \cite{MLCCJCP2014}, a CC2~\cite{CC2CPL1995}-like approach was adopted to define the inactive amplitude equations. The CC2-like equations incorporate $T_1$ amplitude equations to all orders while treating the $T_2$ amplitude equations only up to the first order terms in conjunction with a second-order $[V, T^F ]$ term. By contrast, our approach consistently applies a first-order perturbative treatment to both the $T_1$ and $T_2$ amplitude equations to determine inactive environment CC amplitudes. Similar to the MLCC approach, the computational complexity of the inactive MP-CC amplitude equations is $\mathcal{O}(O^2 V^2 X)$ at the singles and doubles level. However, the future inclusion of $T_3$ amplitudes in the MP-CC framework -- where active non-zero $T_3$ amplitudes induce further relaxation of the environment orbitals -- will not increase the complexity of the low-level projection equation.

In contrast, with the MLCC approach, the complexity of the low-level projection equations would have been higher. However, Koch \textit{et al.}~approximate the triple excitations to include only active indices to reduce the complexity. 
For a more detailed discussion on the differences between MLCC and MP-CC, we refer the interested reader to Appendix~\ref{Sec:complexity_triples}. We have summarized all the comparisons in Table \ref{tab:method_compare}. In that table we use uppercase letters ($I, J, K, L,...$ for holes; $A, B, C, D,...$ for particles)  to denote orbitals in the fragment, and lowercase letters ($i, j, k, l,...$ for holes; $a, b, c, d,...$ for particles) for orbitals in the environment. A side-by-side numerical comparison of MP2-CCSD(II), MLCC and MP-CC is left for future work.


\begin{table}[]
    \centering
    \begin{tabular}{c c c c c}
        method & MP-CCSD & R-CC/PT & MP2-CCSD(II) & MLCCSD  \\
        & (Relaxed) &  &  & \\
        \hline
        reference & this work & \citenum{nooijen1999combining} & \citenum{bochevarov2006hybrid} &  \citenum{MLCCJCP2014}  \\
        \hline
         $T_1^{F}$+ $T_2^{F}$ & All & All & All & All \\
         \hline
         Classes of & All & All & All & All \\
         $T_1^\mathrm{E}$ relaxed & & & & \\ 
         \hline
         Classes of  & All & $\{T^{aB}_{IJ}, T^{AB}_{iJ}$ & $\{T^{aB}_{IJ}, T^{AB}_{iJ}\}$ & All \\
         $T_2^\mathrm{E}$ relaxed & & $,T^{aB}_{iJ}\}$ & & \\
         \hline
         Active space & AVAS & CMO & CMO & CD \\
         basis & & & & \\
         \hline
         Relaxation & Sec.~\ref{sec:h_low} & CCSD & CCSD & CC2 \\
         method & & & & \\ 
         \hline
         Scaling of & $\mathcal{O}(O^2 V^2 X)$ & $\mathcal{O}(O^2 V^3 V_\mathrm{F})$  & $\mathcal{O}(O^2 V^3 V_\mathrm{F})$ & $\mathcal{O}(O^2 V^2 X)$ \\
         relaxation & & & & \\
         method & & & &
    \end{tabular}
    \caption{Comparison of the MP-CCSD embedding scheme with three other related approaches. Note the relaxation method scaling cost is per iteration, and CD stands for Cholesky decomposition based localized orbitals.}
    \label{tab:method_compare}
\end{table}

\section{Implementation}
\label{sec:Implimentation}

The computational results are obtained using a pilot implementation of the MP-CC method at the singles-doubles truncated level, that is MP-CCSD, based on the Python-based Simulations of Chemistry Framework (\texttt{PySCF})~\cite{pyscfJCP2020,sun2018pyscf,sun2015libcint}.
In this implementation, the amplitude equations are separated into different classes for the fragment and environment, allowing us to experiment with the different heuristics mentioned in Sec.~\ref{sec:h_low}. 
We conduct all-electron calculations using CCSD as the computational method in both spin-restricted and spin-unrestricted formulations. For the spin-unrestricted version of MP-CCSD, unrestricted Hartree-Fock (UHF) orbitals will be used.
Since all calculations employ localized orbitals, the MP2 equations, as in Eq.~\eqref{Eq:MP2}, are of the non-canonical type. 

We want to comment on a few aspects of the implementation of steps (4)-(6) in Fig.~\ref{fig:embed_scheme}. We solve the projection equations in both Step 5 and Step 6 iteratively. After solving Step 6 we rebuild $W^F$, and solve Step 5 and 6 again, thereby introducing a global outer loop. The entire set of equations are then solved in a double-loop fashion. Similar to the inner loop that involves step 5 and step 6, we employ the DIIS scheme for the outer loop as well. The number of iterations in the outer loop is usually much smaller than that in the inner loop.

\section{Results} 
\label{sec:results}

\subsection{Diatomic potential energy curves}

We apply our embedding method to the potential energy curves (PECs) of N$_2$ and CO to study the following aspects of the theory: 

\begin{enumerate}
    \item To assess the accuracy of different approaches, we analyze the PECs with different approaches, namely MP2, both relaxed and unrelaxed versions of the MP-CCSD method and also a composite method. For the composite method, we modify Eq.~\eqref{Eq:composite} to make it an uncoupled approximation to the current embedding scheme:

    \begin{equation}
    \begin{aligned}
        E_{\mathrm{Composite}} 
        &= E_{\mathrm{MP2,large}}+E_{\mathrm{CCSD,small(fc)}} 
        \\
        &\quad - E_{\mathrm{MP2,small(fc)}},
    \end{aligned}
    \end{equation}

    where large and small stand for large, small AO basis sets considered for the current calculation, and fc stands for the frozen core calculation (only if it is used).
       
    We compare the error in energy with respect to the CCSD method, using two metrics: the maximum absolute error (MAE) and the non-parallelity error (NPE). The spin unrestricted energy at dissociation will be the zero of energy. We furthermore evaluate several spectroscopic parameters, namely, equilibrium bond distance ($R_e$), harmonic frequency ($\omega_e$), and first anharmonicity constant ($\omega_ex_e$) by the least square fit of the PEC with a polynomial with high enough degree such that the parameter values converge.    
          
    \item We investigate the method's convergence as we systematically increase the size of the active space aiming to assess systematic improvability.
    
    \item We show consistent behavior of the MP-CCSD method with increased AO basis set size when we keep the active space fixed. 
    This observation is significant, considering the challenges often encountered in achieving similar enhancements within many quantum embedding schemes, for example, with Schmidt decomposition-based DMET \cite{dmet_knizia13} and Bootstrap Embedding (BE)-DMET \cite{BE-DMETJPCL19} methods.  
\end{enumerate}

We will employ three different choices for the small basis (MINAO) used for IAO construction, giving three different active spaces as summarized in Table \ref{tab:active_space_n2}. Note that we employ unrestricted orbitals that can break the $ S^2 $ symmetry of the electronic Hamiltonian,
while allowing us to reach lower variational energies than with restricted orbitals. Notably, UHF shows the correct qualitative behavior in the dissociation limit.

\begin{table}[!ht]
    \centering
    \begin{tabular}{c | c | c | c | c}
     & name & active orbitals   & MINAO  & $(n_e,n_\mathrm{orb})$\\
    \hline
    A & minimal & 1s 2s 2p  & STO-3G & (14,10)\\
    B & split valence (SV) & 1s 2s 2p 3s 3p & SV & (14,18)\\
    C & SV + polarization & 1s 2s 2p 3s 3p 3d & cc-pVDZ & (14,28)
    \end{tabular}
    \caption{Choices of active space for the N$_2$ and CO diatomic molecules, and the small basis (denoted as MINAO) used for IAO construction.}
    \label{tab:active_space_n2}
\end{table}

\subsubsection{The nitrogen molecule (N$_2$)}

The PEC of N$_2$ often serves as a benchmark problem in electronic structure theory due to the complexity involved in the dissociation of its triple bond. UCCSD for the whole system will serve as the reference method by which we test the embedded MP-CCSD approximations, employing the aug-cc-pCVDZ and aug-cc-pCVTZ basis sets. While not strictly relevant for assessing embedding approaches, we note that UCCSD is not very accurate relative to a higher rank CC method, UCCSDT. While UCCSD shows correct dissociative behavior, its MAE and NPE versus UCCSDT are~29 mH and~20 mH, respectively with the aug-cc-pCVDZ basis set. 

\begin{figure}
    \centering
    \includegraphics[width=0.47\textwidth]{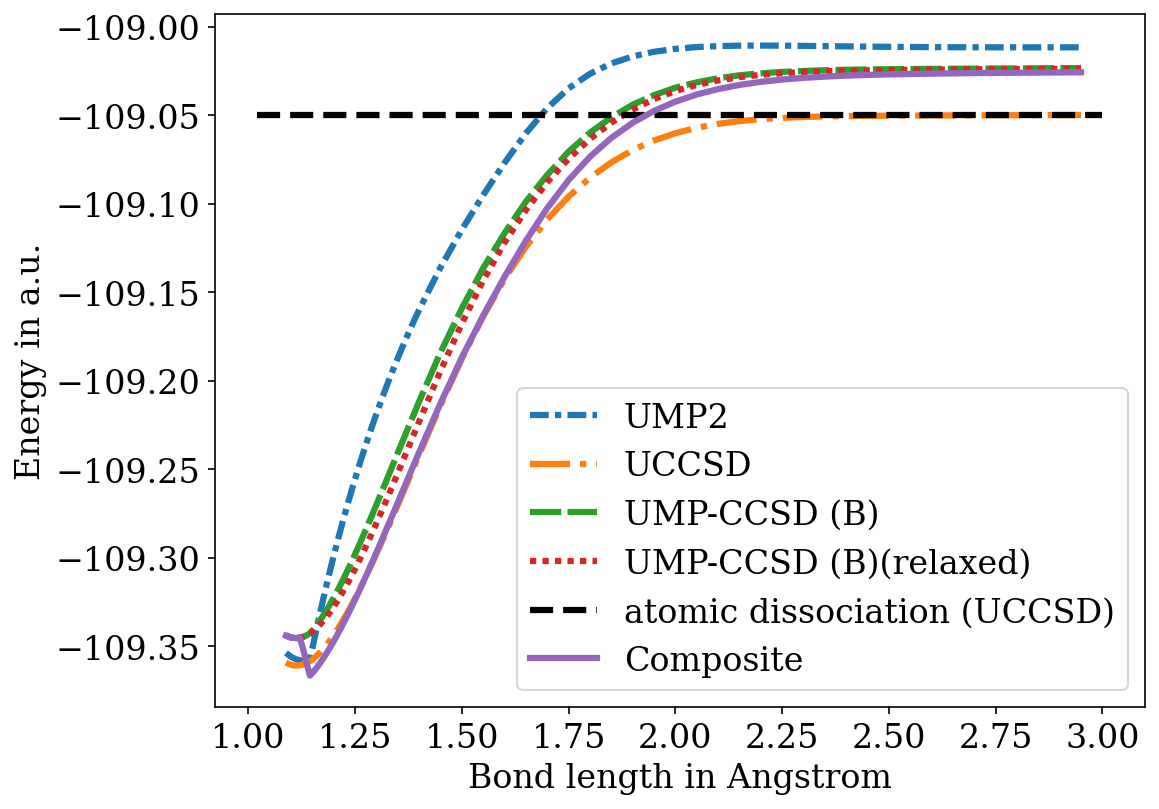}
    \caption{PEC of N$_2$ molecule in aug-cc-pCVDZ basis set with various methods. B active space is as described in Table~\ref{tab:active_space_n2}.}
    \label{fig:abs_ene_n2_3s3p}
\end{figure}

Fig.~\ref{fig:abs_ene_n2_3s3p} shows PECs obtained with the reference UCCSD method and the various approximations, such as UMP2 and the embedding approaches (using the split valence active space B; see Table~\ref{tab:active_space_n2}). We observe that the PECs obtained from various methods are not of similar quality. This is especially true close to the Coulson-Fischer point at 1.2\AA \ where UMP2 and the composite method show a first derivative discontinuity. Its origin lies in the fact that UMP2 is incapable of removing UHF's spin contamination error.\cite{kurlancheek2009violations}
Towards the dissociative limit, all methods are qualitatively satisfactory, due to using unrestricted orbitals.

\begin{figure}
    \centering
    \includegraphics[width=0.47\textwidth]{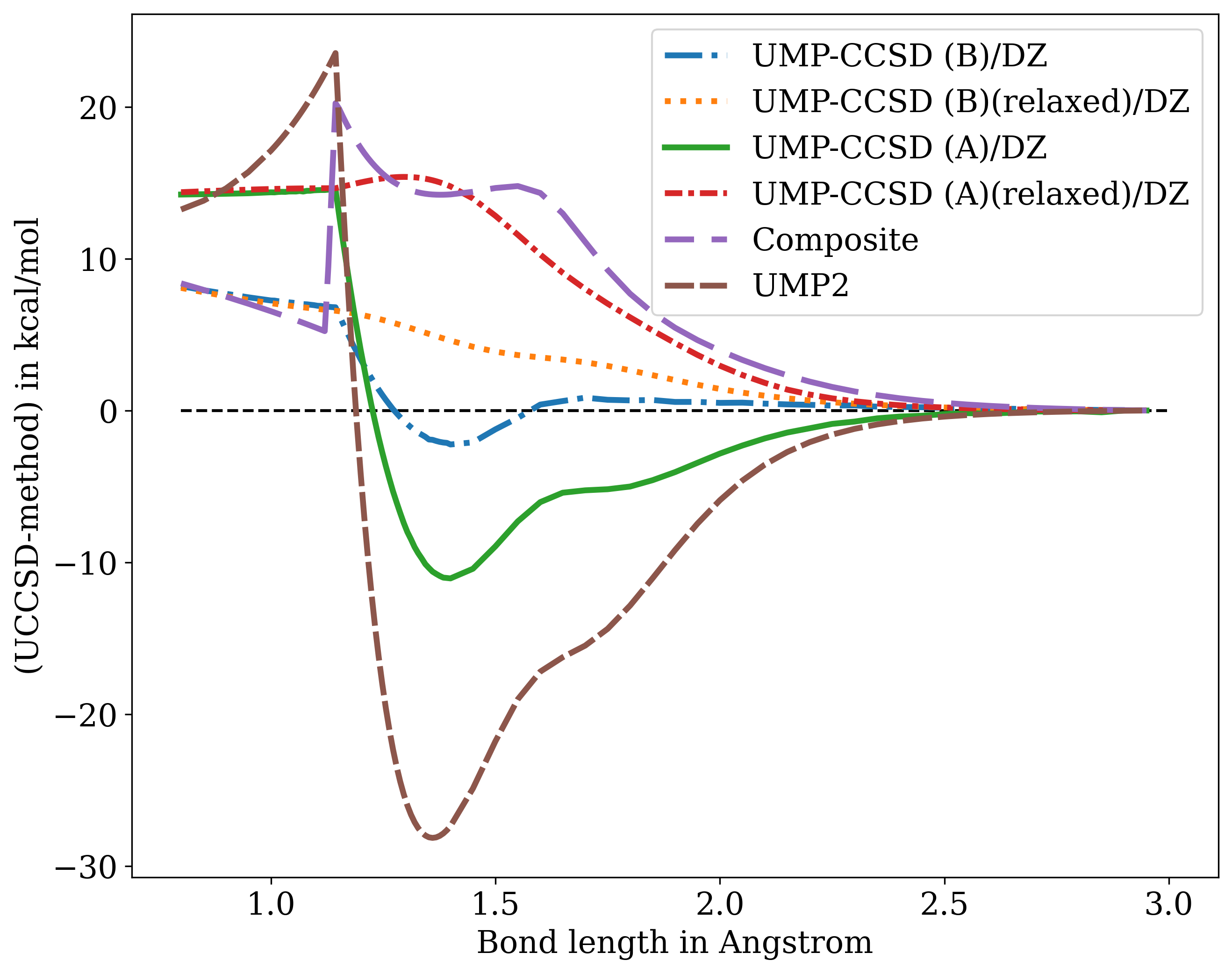}
    \caption{Electronic energy error with respect to UCCSD along a PEC for N$_2$ in the aug-cc-pCVDZ basis set. A and B active spaces are as described in Table~\ref{tab:active_space_n2}. DZ stands for the aug-cc-pCVDZ basis set. We set the energy at dissociation as the zero energy for each method.}
    \label{fig:corr_ene_diff_n2_2s2p3s3p}
\end{figure}

To focus on the relative errors, Fig.~\ref{fig:corr_ene_diff_n2_2s2p3s3p} plots the differences between each approximation and the UCCSD reference, as well as showing embedding results in the minimal active space A. Figure~\ref{fig:corr_ene_diff_n2_2s2p3s3p} shows that UMP2 errors peak near the Coulson-Fischer point and its derivative exhibits discontinuity. The UMP-CCSD method reduces the MAE from~80 mH (UMP2) to~38 mH~/~30 mH, and the NPE from~78 mH (UMP2) to~24 mH~/~17 mH for A~/~B choice of active spaces, respectively. Similarly, the composite scheme reduces MAE and NPE to 24 mH and 24 mH, respectively. However, the composite method contains \textit{two} derivative discontinuities which result from the different Coulson-Fischer points in the small and large basis sets, and is clearly undesirable! The unrelaxed variant of UMP-CCSD shows a derivative discontinuity at the same bond distance as UMP2. Very encouragingly, this discontinuity is lifted using the relaxed version of UMP-CCSD, showing the importance of the \textit{environment singles amplitudes} in symmetry restoration.

\begin{figure}
    \centering
    \includegraphics[width=0.47\textwidth]{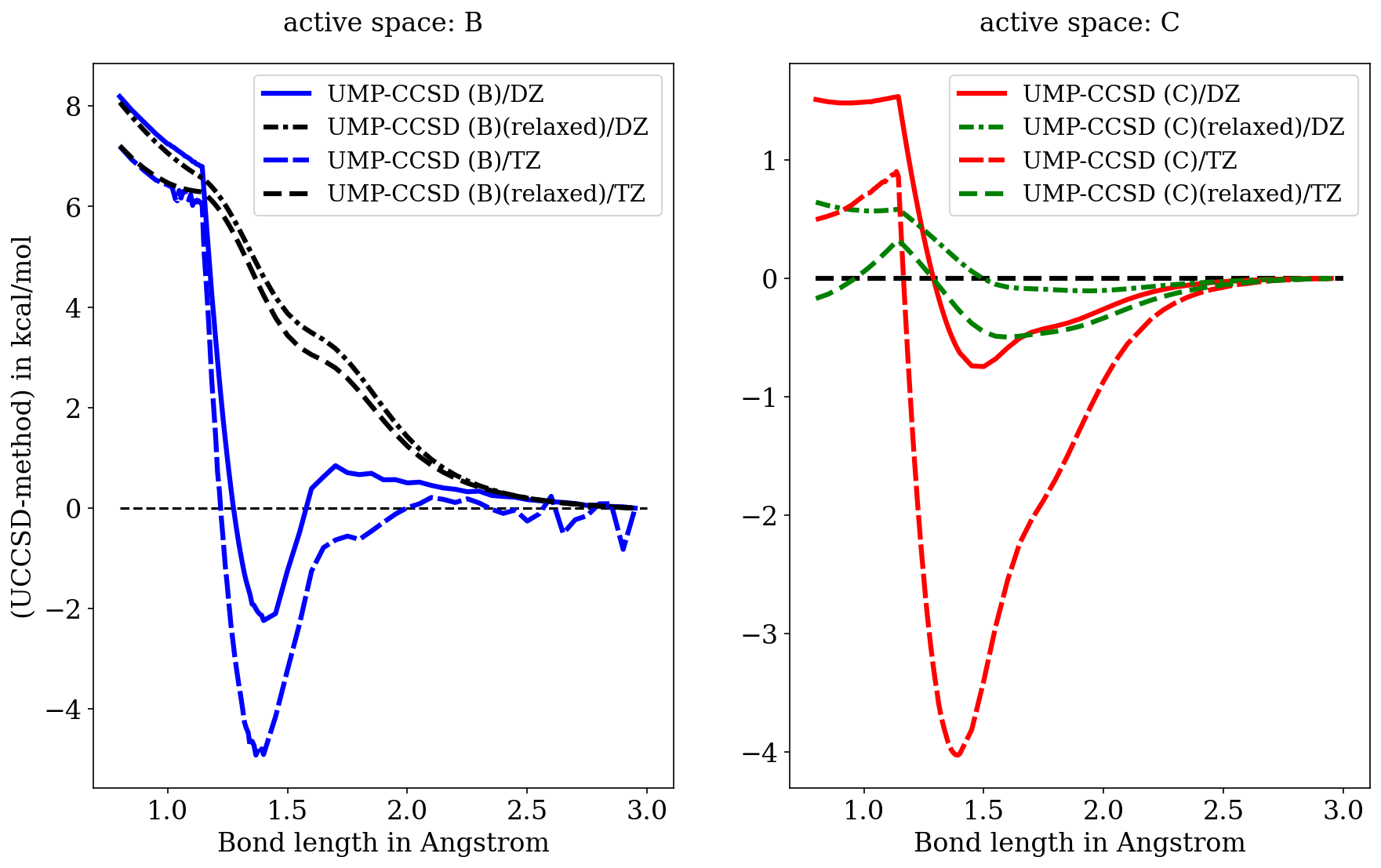}
    \caption{N$_2$ electronic energy error with respect to UCCSD in aug-cc-pCVDZ (denoted as DZ) and aug-cc-pCVTZ (denoted as TZ) basis sets. B and C active spaces are as described in Table~\ref{tab:active_space_n2}. We set the energy at dissociation as the zero energy for each method.}
    \label{fig:corr_ene_diff_n2_3d}
\end{figure}

To further improve the UMP-CCSD energies, we augment the active space with N 3d orbitals (the SVP active space C).
We report the relaxed and unrelaxed UMP-CCSD results in aug-cc-pCVDZ (DZ) and aug-cc-pCVTZ (TZ) basis set for active space choices B and C in Fig.~\ref{fig:corr_ene_diff_n2_3d}. Comparing the left and right panels of Fig.~\ref{fig:corr_ene_diff_n2_3d} shows that MAE and NPE have significantly improved in the SVP active space C compared to the smaller SV active space B. With active space C, the error in energy is well within chemical accuracy ( $<$ 1 kcal/mol) for the relaxed method, and it is close ($\approx$ 2 kcal/mol) with the unrelaxed method.
  
We will now analyze the AO basis set convergence of embedding errors with the aid of Fig.~\ref{fig:corr_ene_diff_n2_3d}. We have enlarged the basis set from aug-cc-pCVDZ (54 basis functions) to aug-cc-pCVTZ (118 basis functions) for the SV (B) and SVP (C) active spaces. For relaxed UMP-CCSD, we observe a nearly constant error as we increase the AO basis from double-$\zeta$ to triple-$\zeta$ quality. However, for the unrelaxed UMP-CCSD variant, we observe that the error increases substantially from double-$\zeta$ to triple-$\zeta$. We also observed a few instabilities in the error curve of the unrelaxed UMP-CCSD method, corresponding to at least two solutions. 
With the relaxed approach, we do not see such instabilities, meaning no evidence of multiple solutions, as well as no loss of accuracy when enlarging the basis set.

The quality of the PECs and the difference curve with various methods, as described so far, influences the spectroscopic parameters as shown in Table~\ref{tab:N2_spec_param}. Compared to UCCSD, we observe that UMP-CCSD reproduces $R_e$, $\omega_e$ and $\omega_e x_e$ better than the composite method and UMP2. The failure of the composite method and UMP2 in estimating $\omega_e$ and $\omega_e x_e$ can be attributed to the discontinuities in the PEC near the equilibrium region. 
Although the relaxed and unrelaxed variants of UMP-CCSD yield similar MAE and NPE along the PEC, the energy difference plots around the equilibrium region become much smoother when orbital relaxations are considered. This improvement is reflected in the spectroscopic parameters $\omega_e$ and $\omega_ex_e$, which are coefficients of higher order terms in the polynomial. They now become significantly closer to the corresponding UCCSD results when orbital relaxations are taken into account. Hence, the self-consistency cycle improves the behavior around the equilibrium region significantly. 

\begin{table}
    \centering
    \begin{tabular}{c|ccc}
       Method & $\Delta R_e$  & $ \Delta \omega_e$  & $ \Delta \omega_ex_e$  \\
        &(in \AA) & (in cm$^{-1}$) & (in cm$^{-1}$)\\
        \hline    
        UMP2 &  0.0059 & 1230.06	& 120.3261  \\
        Composite & 0.0539 & 327.78 & 320.5781 \\
        UMP-CCSD (C) & 0.0003 & 86.85 &	-6.8039 \\
        UMP-CCSD (C)(relaxed) & 0.0006 & 2.60 & -1.7131
    \end{tabular}
    \caption{Spectroscopic parameters of N$_2$ evaluated in aug-cc-pCVTZ basis set. We reported the difference w.r.t.~the UCCSD method for all the spectroscopic parameters. In UMP-CCSD (C), C stands for the active space defined in Table ~\ref{tab:active_space_n2}.}
    \label{tab:N2_spec_param}
\end{table}

\subsubsection{Carbon monoxide (CO)}

\begin{figure}
    \begin{tabular}{c }

    \includegraphics[width=0.45\textwidth]{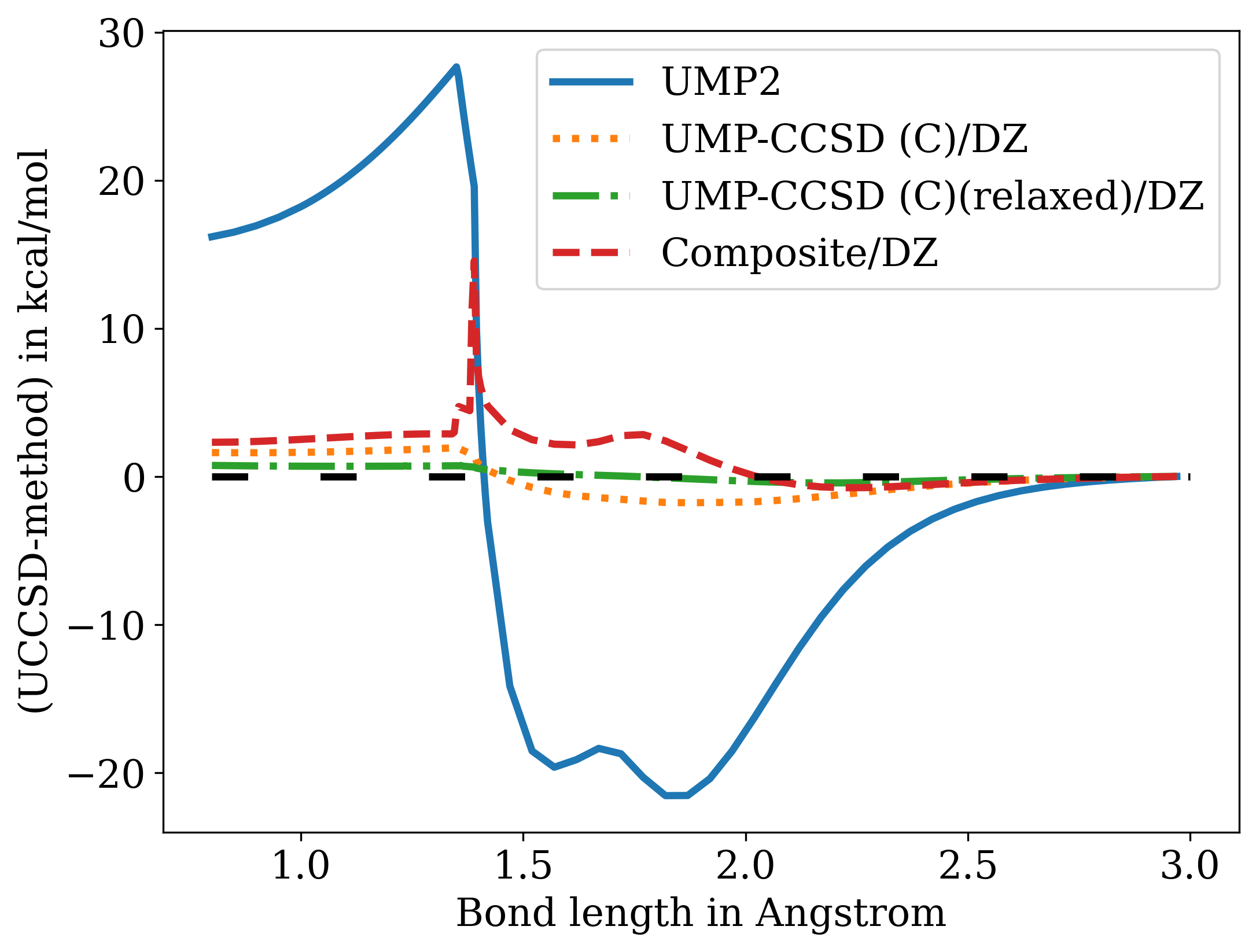} \\
    \includegraphics[width=0.45\textwidth]{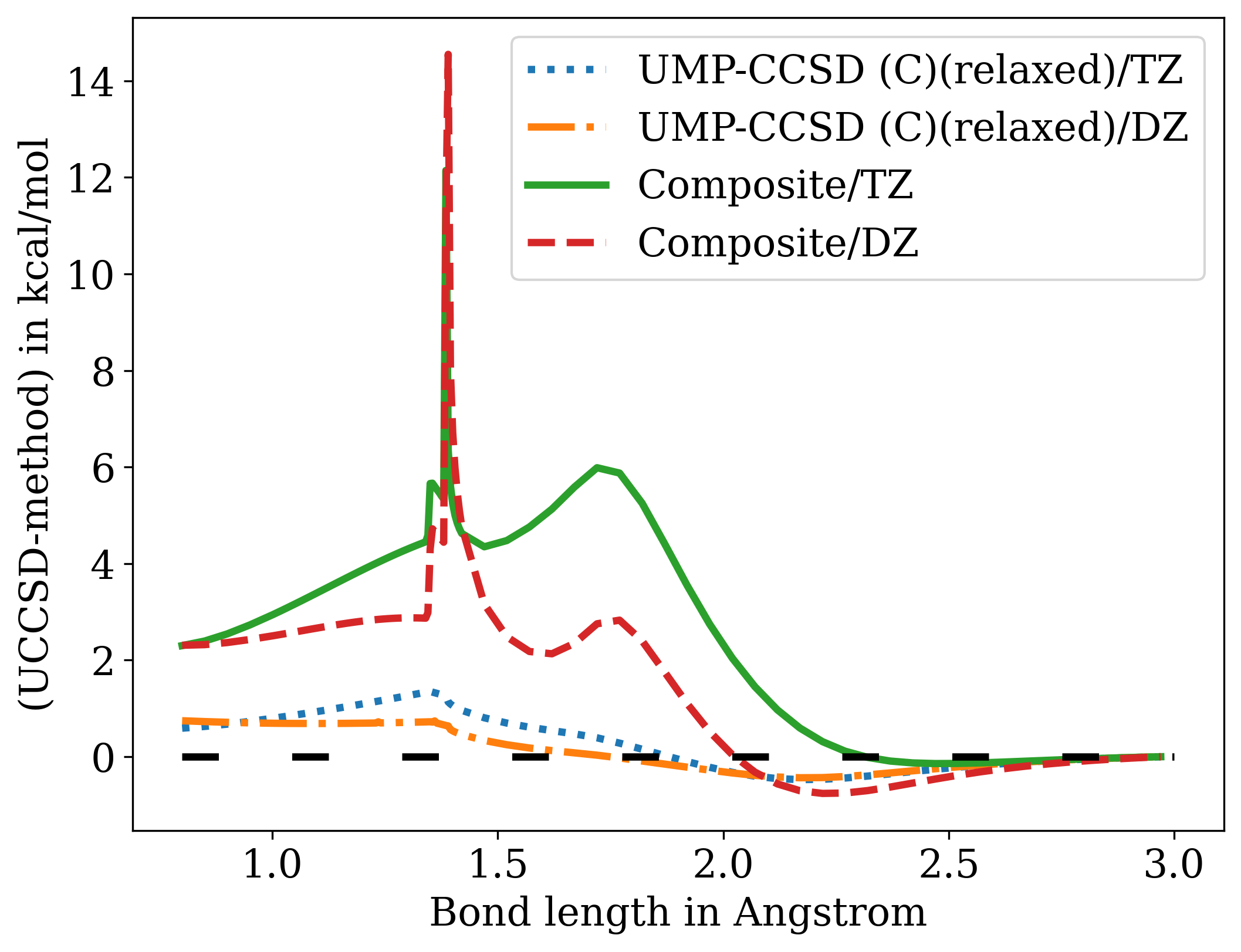}
         
    \end{tabular}
    \caption{Electronic energy differences w.r.t.~UCCSD along a PEC for CO. DZ and TZ are abbreviations for aug-cc-pCVDZ and aug-cc-pCVTZ basis sets respectively. We set the energy at dissociation as the zero energy for each method.}
    \label{fig:corr_energy_diff_CO}
\end{figure}

We now consider the isoelectronic carbon monoxide molecule. The full PECs for UMP2, UCCSD, relaxed and unrelaxed UMP-CCSD, and the composite method are given in Appendix Fig.~\ref{fig:pec_co}. Beginning with the cc-pVDZ basis and the large SVP active space C, the energy difference curves for all methods are shown in Fig.~\ref{fig:corr_energy_diff_CO} (top panel). There are two derivative discontinuities in the UMP2 curve and multiple derivative discontinuities in the composite method curve. 
The latter arises because the composite method inherits discontinuities from its individual computations; here multiple discontinuities appear at different bond distances for different basis sets, see ``UMP2 (large)'' and ``UMP2 (small)'' graphs in Appendix Fig.~\ref{fig:corr_energy_diff_composite_co}.
Using the unrelaxed version of UMP-CCSD, we observe one derivative discontinuity, which can be lifted by including orbital relaxations.   

The bottom panel of Fig.~\ref{fig:corr_energy_diff_CO}, plots the basis set dependence of the error in the relaxed embedding method and the composite method. As for N$_2$, we observe less than 1 kcal/mol loss of accuracy when enlarging the basis set from aug-cc-pCVDZ to aug-cc-pCVTZ. 
For the uncoupled composite scheme, the extent of agreement varies strongly across the PEC, and is qualitatively worse than the relaxed UMP-CCSD method. This issues the value of electronic embedding.

We also calculate the spectroscopic parameters $R_e$, $\omega_e$ and $\omega_e x_e$ for the carbon monoxide molecule, which are listed in Table~\ref{tab:CO_spec_param}. Compared to the preceding N$_2$ example, the composite method shows noticeable improvement over UMP2 for all computed quantities. We attribute this to the qualitative agreement of the composite method's PEC with the PEC of UCCSD near the equilibrium region, see Appendix Fig.~\ref{fig:pec_co}, and the fact that the discontinuities appear further away from equilibrium. We still observe that $R_e$ and $\omega_e$ are best reproduced with the UMP-CCSD method, showing a consistent improvement over the composite method.

\begin{table}
    \centering
    \begin{tabular}{c|ccc}
        Method & $\Delta R_e$  & $\Delta \omega_e$  &  $\Delta \omega_ex_e$  \\
        &(in \AA) & (in cm$^{-1}$) & (in cm$^{-1}$)\\
        \hline
        UMP2 & 0.0101 & -105.82 &	1.221 \\
        Composite & 0.0017 &	-12.76 & -0.0878 \\
        UMP-CCSD (C) & 0.0007 & -6.73 & -0.1200 \\
        UMP-CCSD (C)(relaxed) & 0.0005 & -4.65 & 0.0817
    \end{tabular}
    \caption{Spectroscopic parameters of CO evaluated in aug-cc-pCVTZ basis set. We reported the difference to the UCCSD results for all spectroscopic parameters. Note that ``(C)'' in ``UMP-CCSD (C)'', stands for the active choice defined in Table ~\ref{tab:active_space_n2}.}
    \label{tab:CO_spec_param}
\end{table}

\subsection{W4 Dataset}

We now investigate the accuracy of the embedding approaches on thermochemical energy differences using the well-known W4-11 \cite{W4_KARTON_CPL2011} dataset. The cc-pCVTZ basis set is used. We have chosen cc-pVDZ as the small basis reference to generate the fragment active space (i.e., like active space C in Table \ref{tab:active_space_n2}). All atoms are again included in the fragment.
We use only the W4-11 single reference (SR) systems, for which UCCSD serves as the reference to test the MP-CCSD embedding models as well as the composite model and MP2. The multi-reference (MR) systems are less accurately described than the SR systems via UCCSD~\cite{JLeeAFQMC_JCTC22} and are excluded. Altogether, we compute 140 total atomization energies (TAEs) and bond dissociation energies (BDEs), 20 isomerization energies (ISO), 505 heavy atom transfer (HAT), and 13 nucleophilic substitution (SN) reactions. 

\begin{figure}
  \begin{tabular}{c c}
    \includegraphics[width=0.25\textwidth]{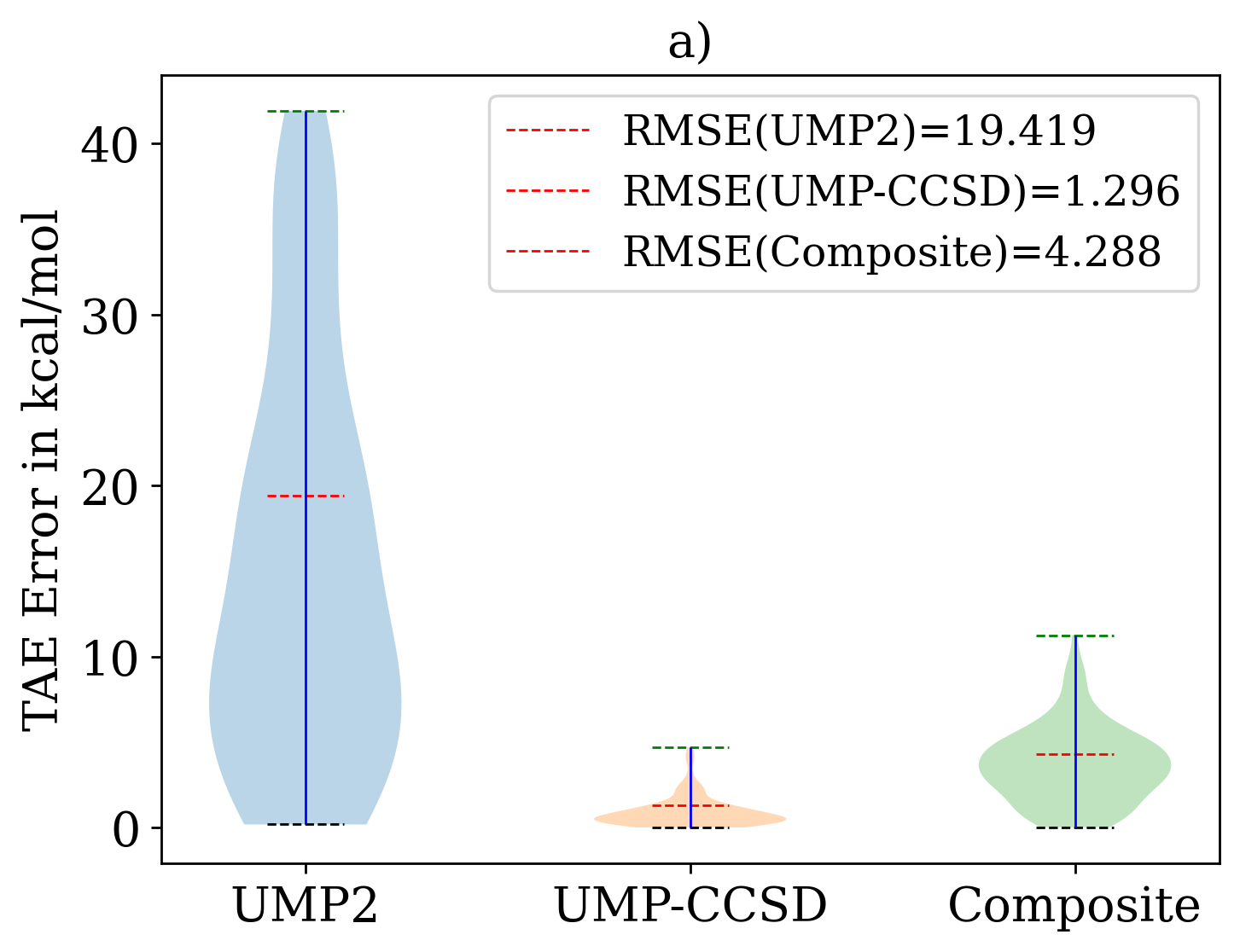}  &   \includegraphics[width=0.25\textwidth]{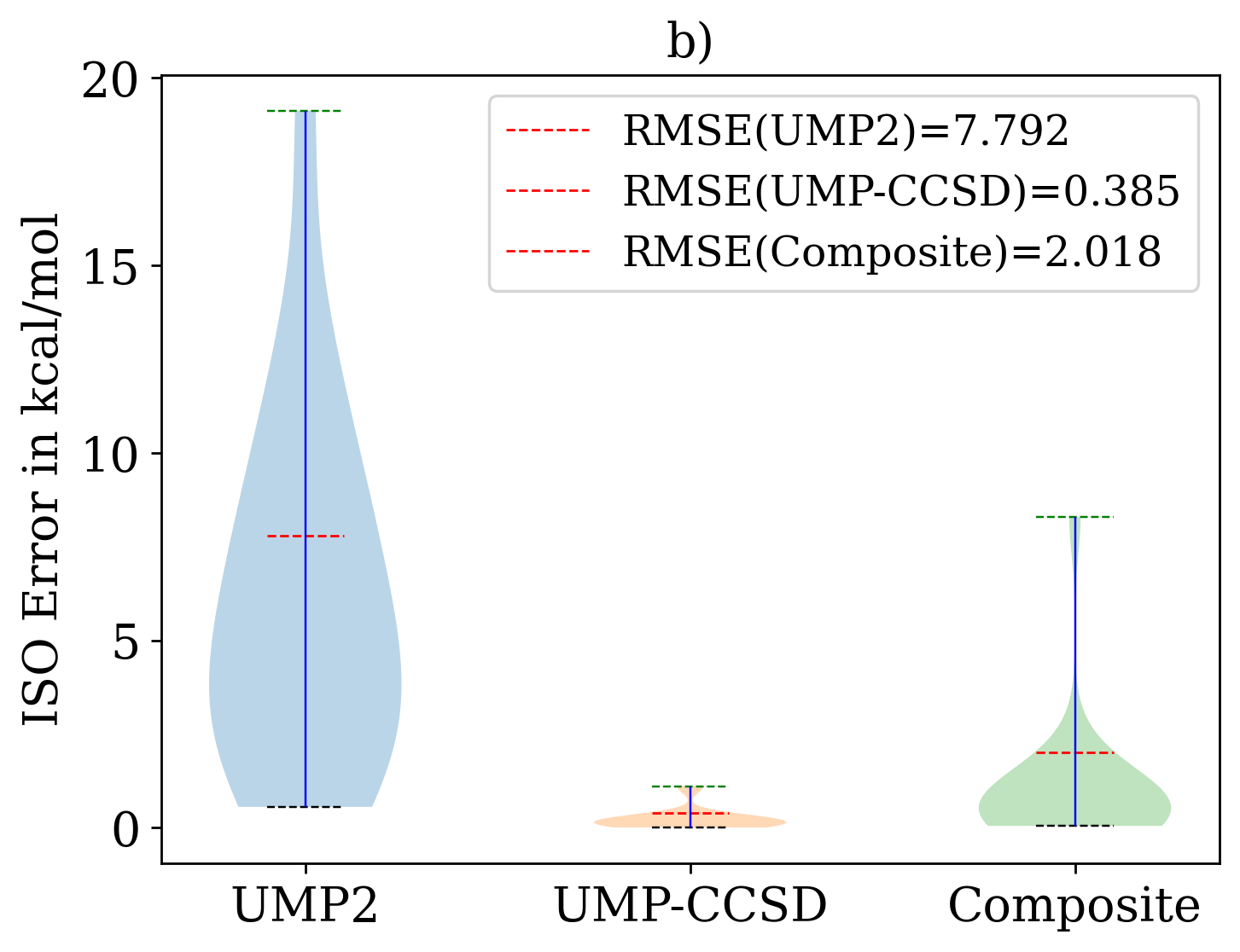} \\
    \includegraphics[width=0.25\textwidth]{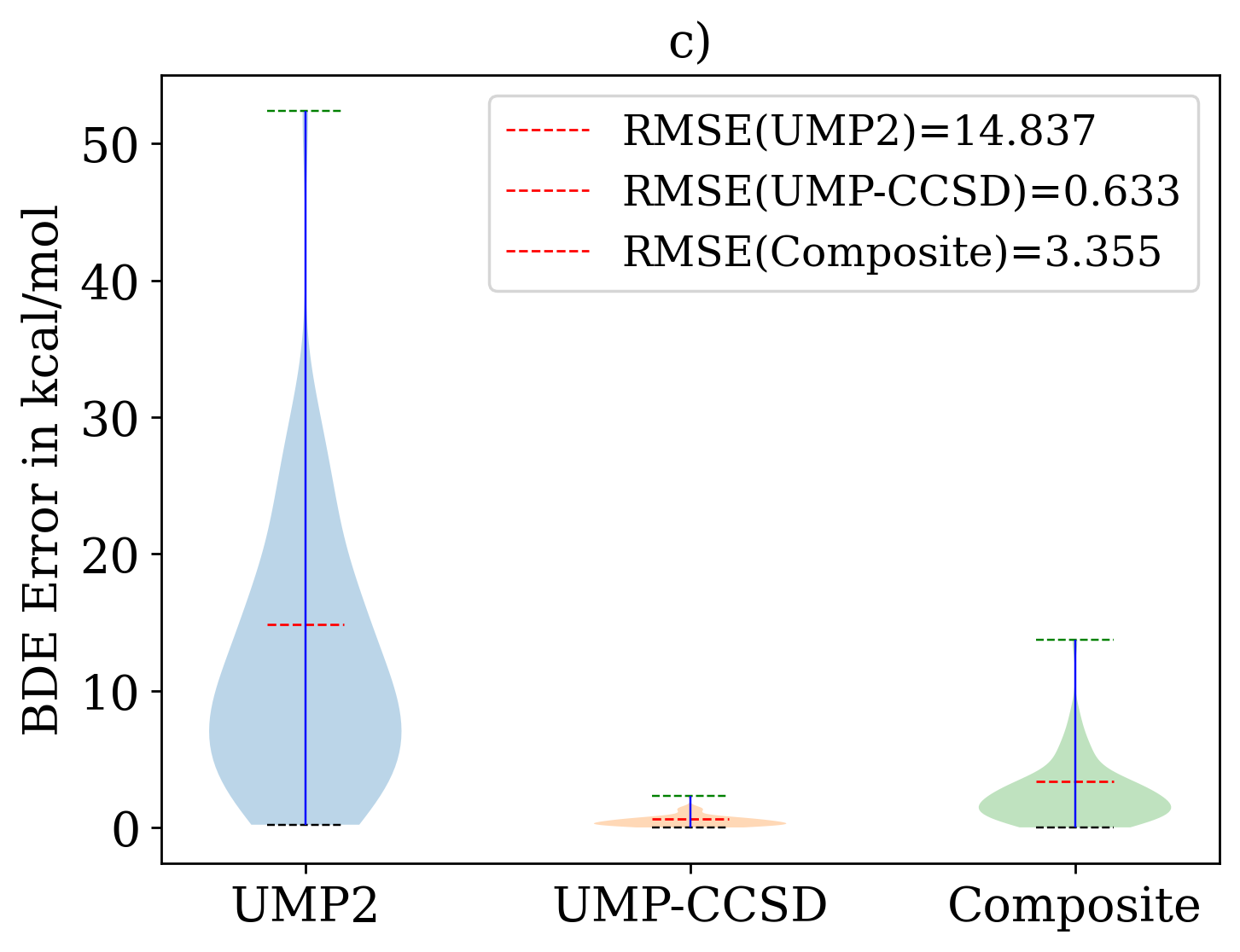}   &  \includegraphics[width=0.25\textwidth]{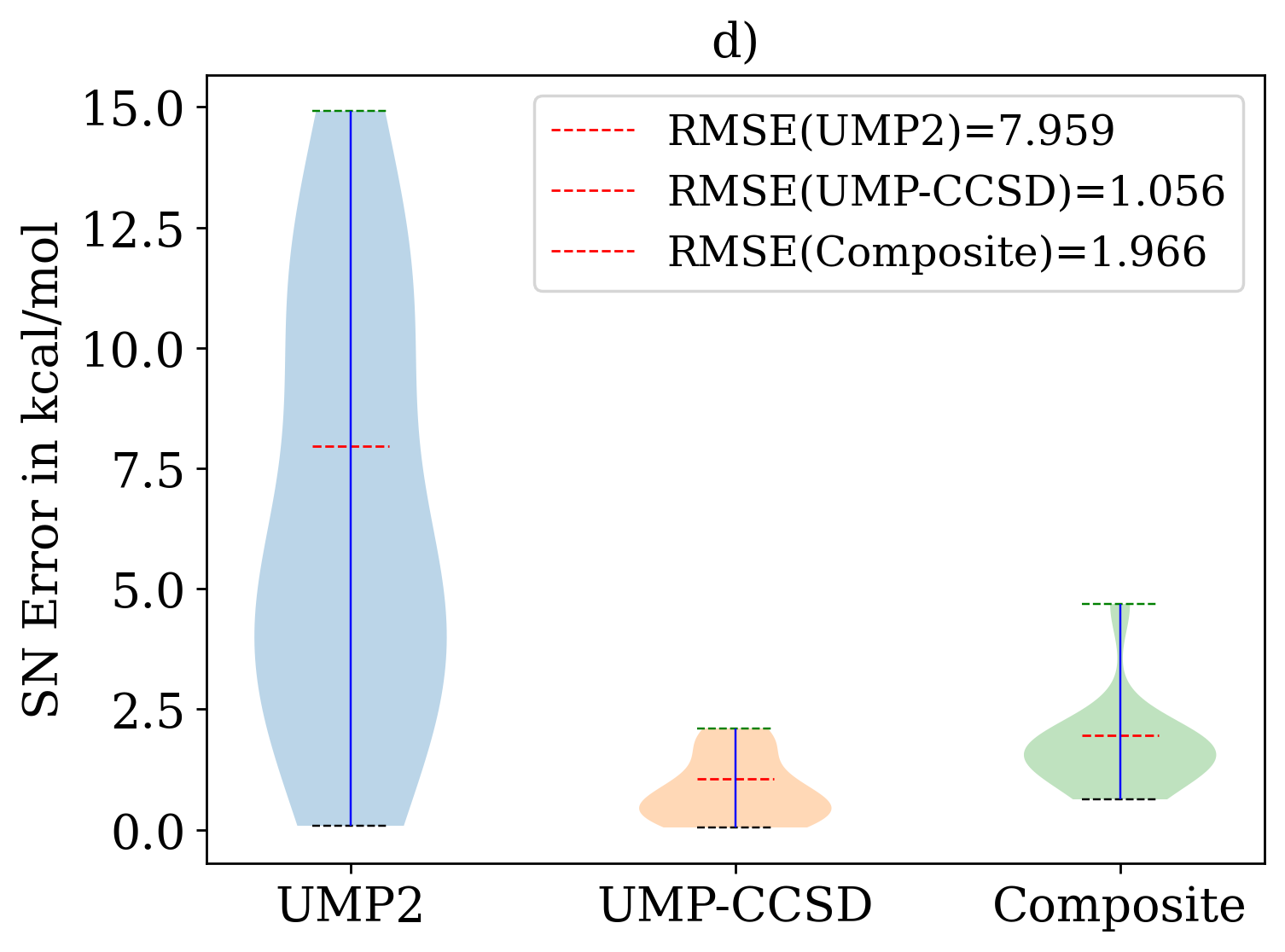} \\
    \includegraphics[width=0.25\textwidth]{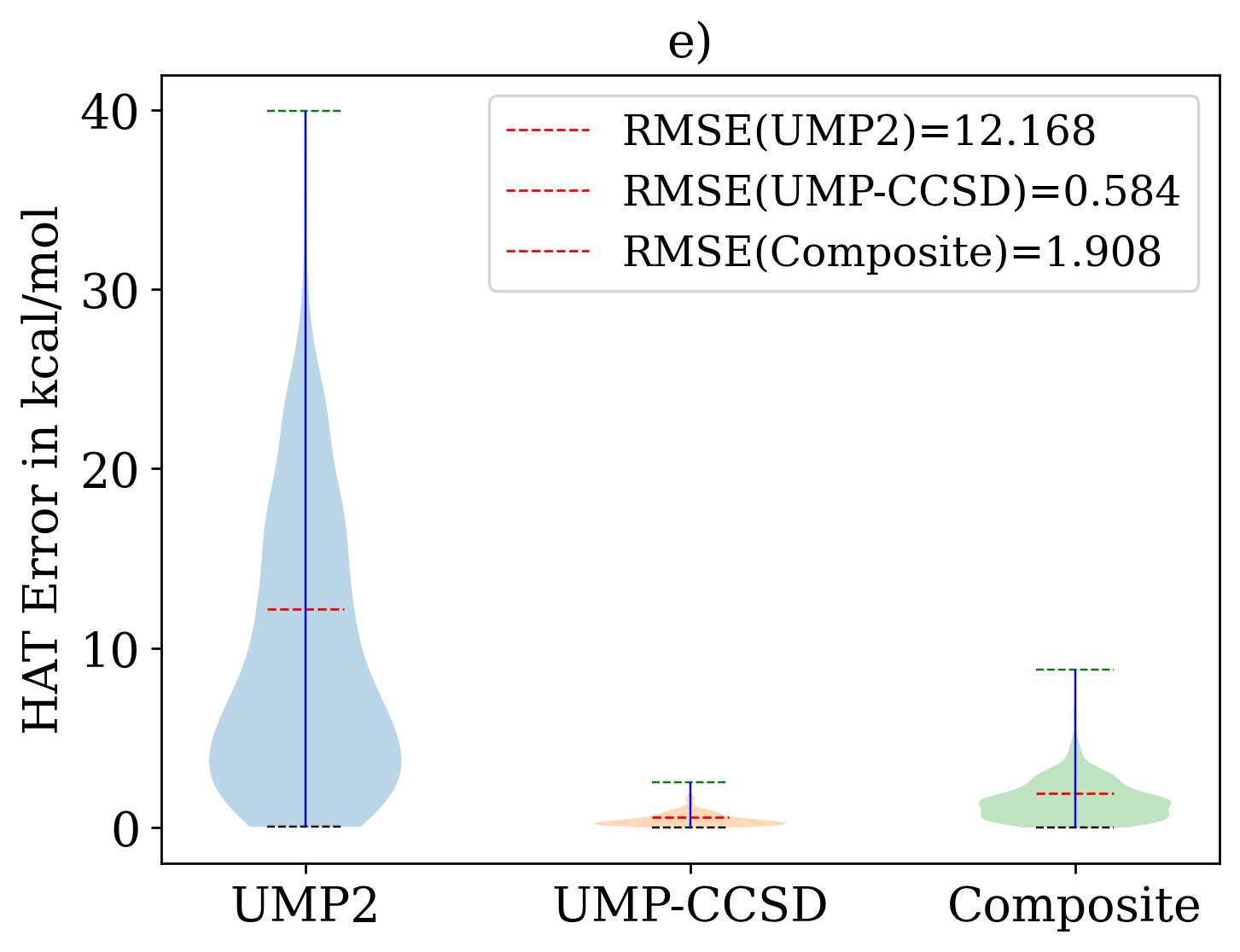} &
  \end{tabular}

    \caption{Violin plot of the errors of UMP2, relaxed UMP-CCSD, and the composite method compared to the corresponding CCSD results for various thermochemical reactions with W4-11 dataset in the cc-pCVTZ basis set. The different panels show a) the total atomization energy, b) the isomerization energy, c) the bond dissociation energy, d) the nucleophilic substitution reaction energy, and e) the heavy atom transfer energy. All atoms are included in the fragment, with active orbitals defined via the use of cc-pVDZ in the AVAS scheme.}
    \label{fig:w4_dataset}
\end{figure}

Figure~\ref{fig:w4_dataset} shows violin plots of the absolute errors relative to UCCSD for different thermochemical reactions. 
We observe that the root mean square error (RMSE) in UMP-CCSD is significantly improved compared to UMP2 for all classes of reactions. Moreover, relaxed UMP-CCSD provides a significant improvement over the composite results. It is encouraging that the RMSE of UMP-CCSD is at or within chemical accuracy for all the reactions considered in this study. Additionally, we observe that the maximal spread of the deviation is also much more favorable for relaxed UMP-CCSD than for the other methods considered.

\subsection{Rotational isomers of 1,3-butadiene}

\begin{figure}
  \begin{tabular}{c c c}
    \includegraphics[width=0.15\textwidth]{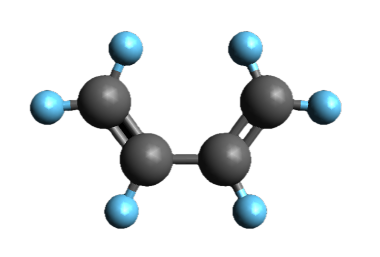}  & \includegraphics[width=0.15\textwidth]{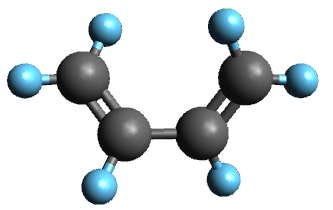}  & \includegraphics[width=0.15\textwidth]{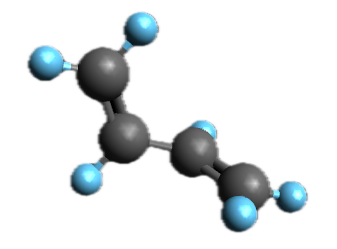} \\
     $\tau = 0 ^\circ $ & $\tau = 35.5 ^\circ $ & $\tau = 90 ^\circ$ \\
      \includegraphics[width=0.15\textwidth]{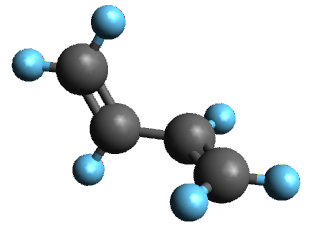}  & \includegraphics[width=0.15\textwidth]{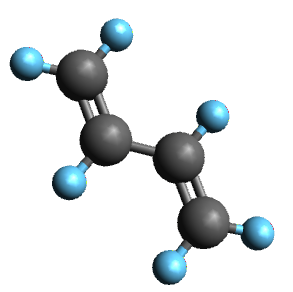} & \\
     $\tau = 101.7 ^\circ $ & $\tau = 180 ^\circ $ & \\
  \end{tabular}
    \caption{Rotational isomers of 1,3-butadiene.}
    \label{fig:rotamer_butadiene}
\end{figure}

In this section, we report on the torsional potential around the central \ce{C-C} bond of 1,3-butadiene, \ce{CH2=CH-CH=CH2} This conjugated molecule has been extensively studied in the literature,\cite{head1993internal, butadiene_FellerJPCA09, butadiene_karpfken_MolPhys04} as its torsional potential is an archetype of the effect of loss of conjugation between the two double bonds at torsional angles ($\tau$) close to $90^\circ$. More specifically, we have studied five different rotamers as shown in Fig.~\ref{fig:rotamer_butadiene}. The optimized structures for all those isomers were taken from Feller \textit{et al.}\cite{butadiene_FellerJPCA09}, optimized at the highly accurate CCSD(T)/aug-cc-pVQZ level of theory. Following our previous examples, we have compared the UMP-CCSD and UMP2 methods with the all-electron UCCSD method, using the cc-pCVTZ basis set. Unlike the previous examples, we have chosen a fragment that is localized spatially as well as limited in its active orbitals. Thus we use the minimal STO-3G basis on only the 4 carbon atoms and consider only the C 2s 2p orbitals to comprise the active space. While the molecule has 30 electrons, 22 of them remain in the fragment, which has a total of 25 orbitals. To construct IAOs within the AVAS scheme, we have considered the STO-3G basis set as our minimal basis.

\begin{figure}
    \centering
    \includegraphics[width=0.5\textwidth]{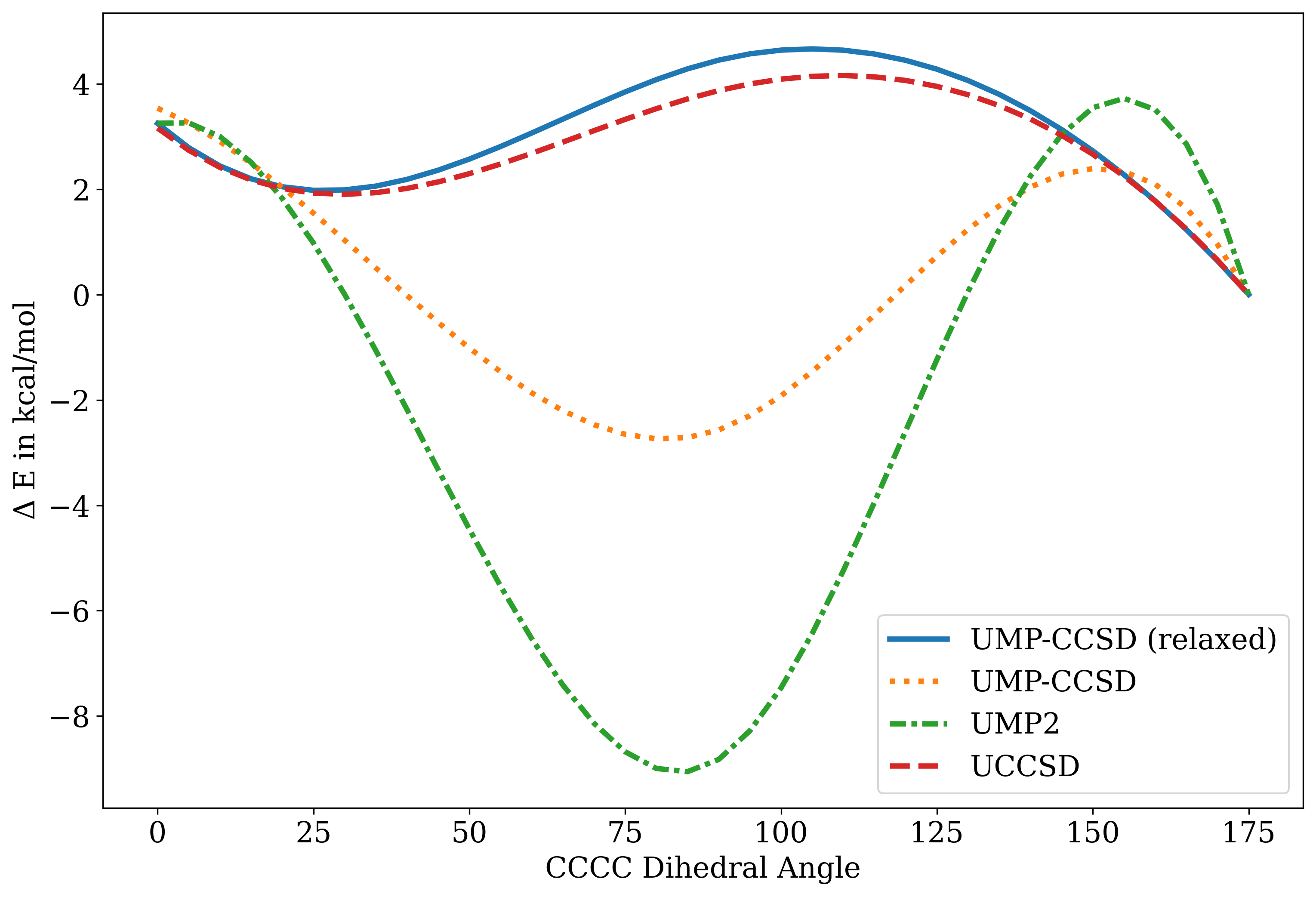}
    \caption{Relative energy of different rotational isomers along the torsional potential around the central \ce{C-C} bond of the butadiene molecule, \ce{H2C=CH-CH=CH2}, with various unrestricted methods. The \textit{trans}, that is, the isomer at $\tau = 180^{\circ}$ has been considered as the zeroth of energy. See text for computational details.}
    \label{fig:rot_butadiene}
\end{figure}

While it is not widely known (and indeed we were surprised to observe it), HF results exhibit spin-polarization all along the torsional potential (see Appendix Table \ref{tab:s2_butadiene}), so UHF energies lie below RHF energies. This indicates the presence of some biradicaloid character in butadiene, which can be interpreted as arising from the \ce{^{.}CH2-CH=CH-^{.}CH2} resonance structure. It is well-known that UMP2 cannot recover symmetries that are broken at the UHF level, due to the absence of orbital rotations\cite{nobes1987slow}, sometimes leading to poor results.\cite{JLEEMHG_OOMP2_JCP18} Indeed this artificial HF symmetry-breaking turns the torsional potential of butadiene from an easy problem for RMP2\cite{head1993internal, butadiene_karpfken_MolPhys04} into a challenging problem for UMP2. This is evident in Fig.~\ref{fig:rot_butadiene} where the relative energy of all the isomers was plotted relative to the \textit{trans} isomer.  UMP2 inverts the correct form of the torsional potential, predicting  $\tau = 90^{\circ}$ as the most stable structure, whereas it should be very close to the barrier.

\begin{figure}
    \centering
    \includegraphics[width=0.5\textwidth]{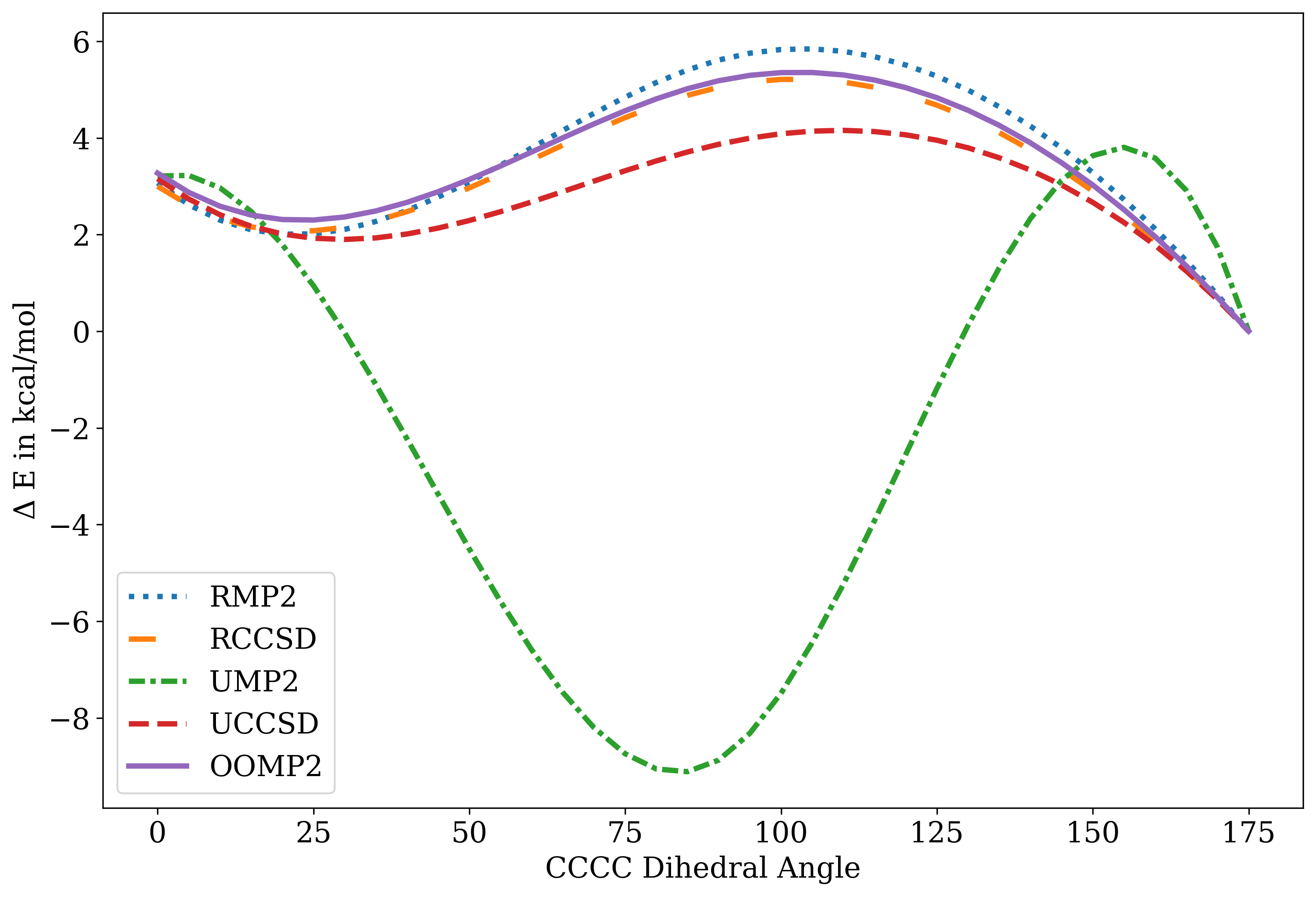}
    \caption{Relative energy of different rotational isomers of the butadiene molecule (w.r.t. the trans isomer) with both restricted and unrestricted methods. See text for computational details.}
    \label{fig:res_un_butadiene}
\end{figure}

By contrast, UCCSD relative energies agree quite well in comparison to the benchmark theoretical data reported by Feller \textit{et al.}\cite{butadiene_FellerJPCA09}, due to the restoration of spin symmetry breaking to a large extent (see Appendix Table \ref{tab:s2_butadiene}) in the presence of singles amplitudes. Turning to the UMP-CCSD methods, due to the absence of inactive singles in the unrelaxed version of the UMP-CCSD method, it is still qualitatively incorrect, although improved relative to UMP2. By contrast, upon inclusion of orbital rotations via $T_1$ the relaxed version of the UMP-CCSD method improves significantly upon both the UMP2 method and the unrelaxed UMP-CCSD method and shows results that are qualitatively and quantitatively very similar to the UCCSD method. The interpretation is straightforward: $T_1$ has largely removed the symmetry-breaking seen at the UHF level. This is further evidenced in Fig.~\ref{fig:res_un_butadiene}, where the orbital-optimized MP2 (OOMP2) \cite{JLEEMHG_OOMP2_JCP18} method based on an unrestricted reference also removes the inverted trend of the UMP2 method.


\subsection{Azomethane}

In order to further test the efficiency of the UMP-CCSD method in capturing system relative energies using a spatially localized active fragment we have studied the PEC of azomethane, \ce{H3C-N=N-CH3}, along the N=N double bond stretching coordinate. All the geometric parameters, except the N=N bond distance of this molecule, were taken from Hermes \textit{et. al} \cite{Hermes_LASSCF_JCTC19}. We use the aug-cc-pCVDZ basis set, with all electrons correlated. For the active fragment, we have chosen only the 2s and 2p orbitals of the 2 nitrogen atoms, as obtained via the AVAS scheme. Also, to construct the IAOs within the AVAS scheme, we have chosen the minimal STO-3G basis set, resulting in a very small fragment consisting of 16 electrons in 16 orbitals.

\begin{figure}
    \centering
    \includegraphics[width=0.5\textwidth]{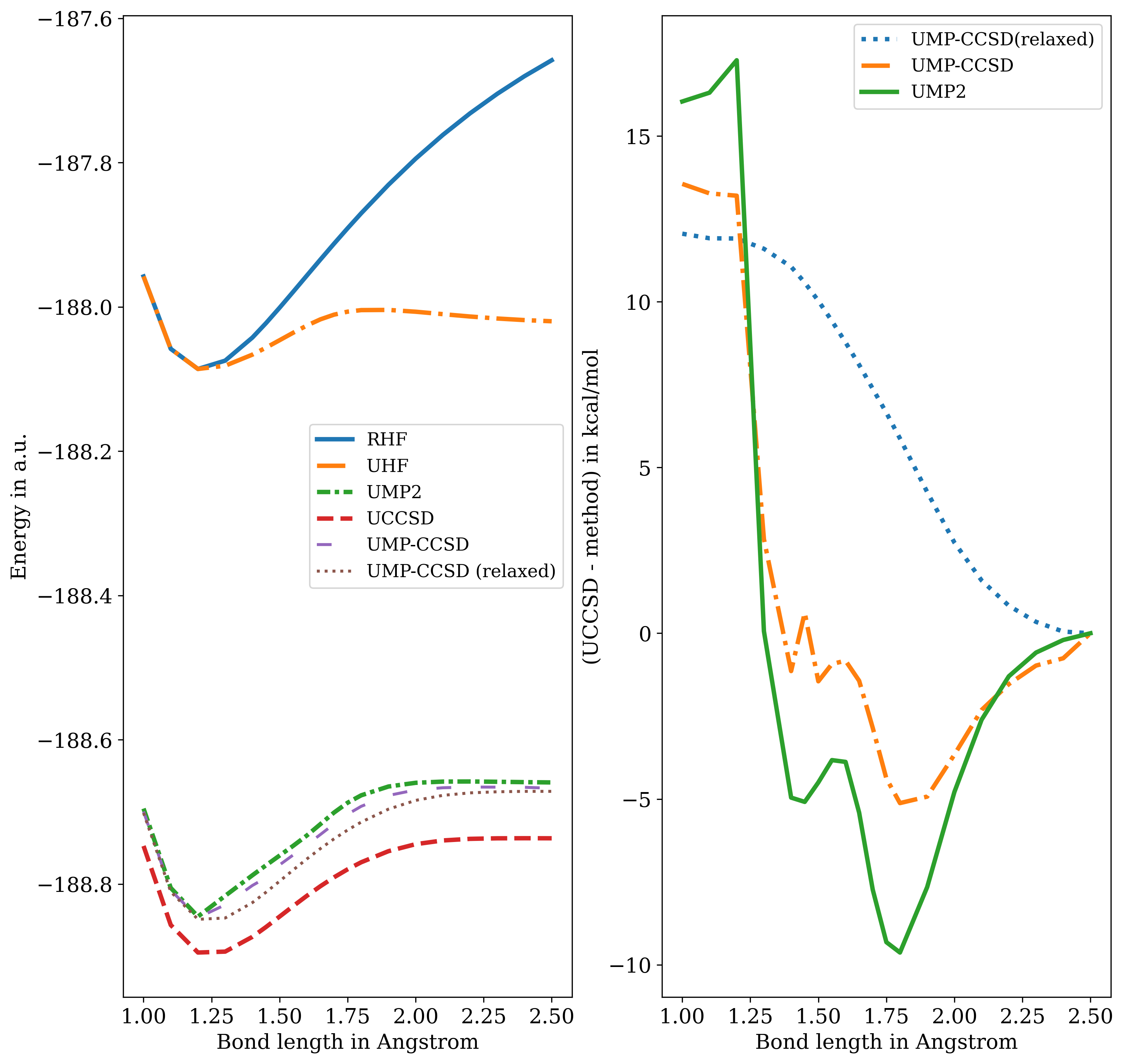}
    \caption{ (left panel) PEC of N=N double bond stretching of azomethane in the cc-pCVDZ basis set. (right panel) Difference plot w.r.t. the UCCSD method. The smallest (valence) active space on the 2 N atoms, as described in the text, is used for this plot. For the difference plot, we set the energy at dissociation as the zero energy for each method}
    \label{fig:pec_azomethane}
\end{figure}

In Fig.~\ref{fig:pec_azomethane} we have plotted the PEC with various methods and also the difference plot with respect to the UCCSD method. First we identify the Coulson-Fischer (CF) point by plotting the PEC with the RHF and UHF methods, which separate at that point. The UMP2 method shows a pronounced first derivative discontinuity (kink) at the CF geometry. The unrelaxed version of the UMP-CCSD method numerically improves upon the UMP2 method, but it cannot remove the kink. However, the relaxed UMP-CCSD method produces a smooth PEC, illustrating the key role of environment $T_1$ amplitudes. 

These characteristics of the PEC w.r.t.~the UCCSD method are even more pronounced in the difference plot (right panel of Fig.~\ref{fig:pec_azomethane}). Interestingly, we observe apparent derivative discontinuities at bond lengths longer than the CF point with both UMP2 and unrelaxed UMP-CCSD, suggesting that the UHF solution changes character there. The relaxed version of the UMP-CCSD method, on the other hand, removes all discontinuities very satisfactorily. This analysis suggests a qualitatively correct behavior of the relaxed version of the UMP-CCSD method for local chemistry. However, the NPE of the PEC remains quite high with the current (very small) choice of active space, which presumably is coming from the larger error towards the dissociation regime. 

\begin{figure}
    \centering
    \includegraphics[width=0.5\textwidth]{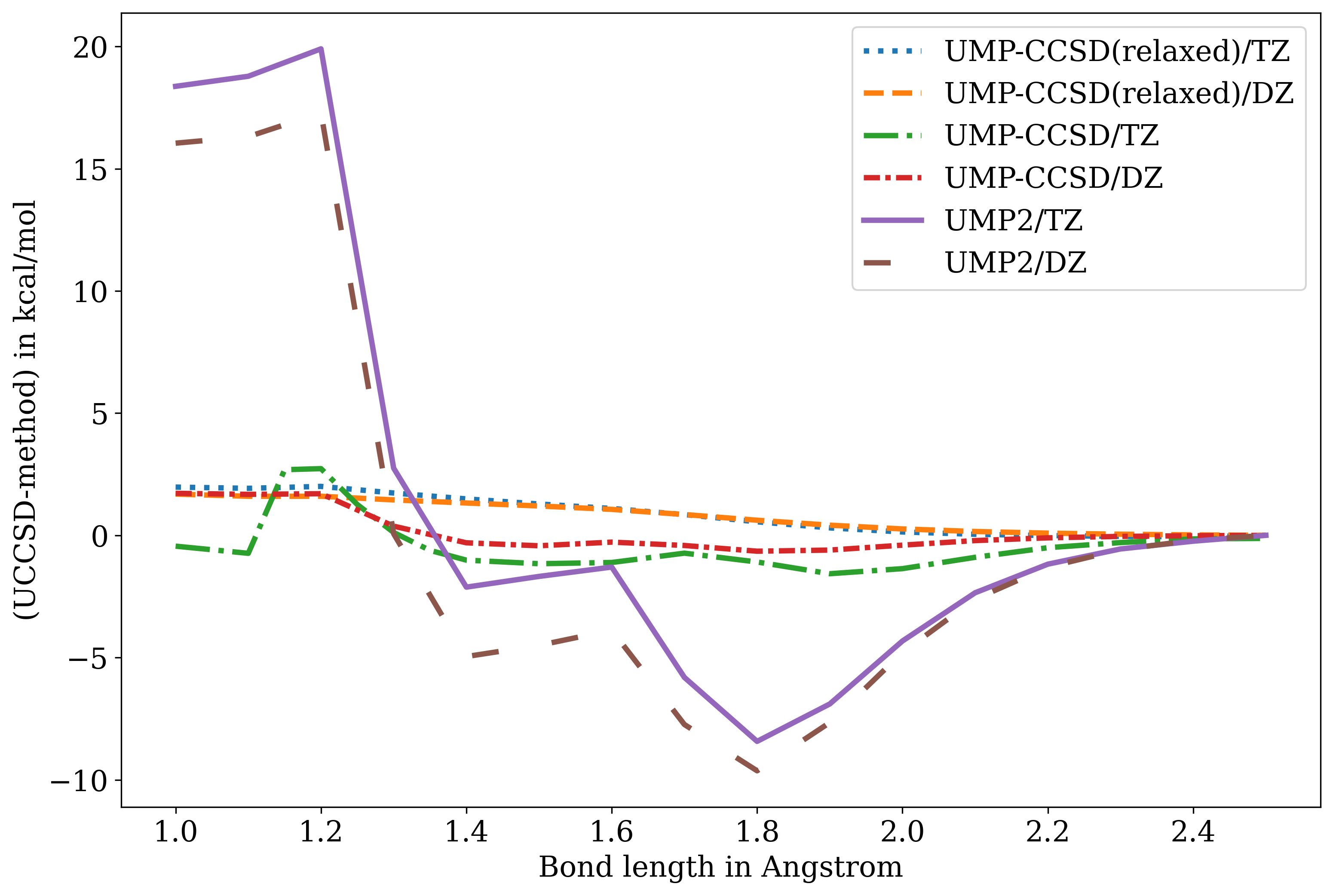}
    \caption{Energy difference w.r.t. the UCCSD method for azomethane molecule. DZ and TZ stand for the cc-pCVDZ and cc-pCVTZ basis sets, respectively. A larger local active space on the 2 N atoms, as described in the text, is used for this plot. We set the energy at dissociation as the zero energy for each method.}
    \label{fig:pec_azomethane_tz}
\end{figure}

To reduce the NPE, we expanded the size of the active space, by using N 2s 2p 3s 3p 3d orbitals, where cc-pVDZ basis set was considered as the reference basis set for IAO construction. Fig.~\ref{fig:pec_azomethane_tz} shows that this choice significantly reduces the NPE. We then further increased the size of the basis set for the actual calculation to cc-pCVTZ, and chose the latter (larger) active space. We observed that when the relaxed scheme is employed, the error remains almost constant w.r.t. the cc-pCVDZ basis set, which corroborates our previous observations for diatomic dissociation.


\section{Conclusion}
 
We have proposed and numerically scrutinized the new MP-CC quantum embedding scheme, where the active fragment problem is solved at the coupled cluster (CC) level, and the environment problem is solved at a level similar to 2nd order M{\o}ller--Plesset (MP) perturbation theory. 
Instead of constructing a density matrix or Green's function at the CC level, MP-CC relies only on the cluster amplitudes to separate the environment and the fragment problem. 
Self-consistency is ensured through amplitude equations coupling the fragment and the environment. 
In particular, we do not need to construct a hybridization term between these two layers to take ``environment fluctuations'' into account. 
Furthermore, the interaction term for the fragment problem is renormalized by the low-level environment amplitudes.

MP-CC provides a general framework allowing the use of different low-level methods for solving the environment problem, while the high-level method for the fragment problem remains fixed as a CC method. 
In this work, we fixed the CC method as CCSD, and proposed and compared two different MP2-like environment models: one without orbital relaxation and one with orbital relaxation.
The relaxed version significantly improves upon the unrelaxed one by removing a derivative discontinuity in the error curve relative to UCCSD results
By contrast, a composite method (or the uncoupled embedding method) numerically improves upon UMP2, but inherits first derivative discontinuities of UMP2, making it significantly inferior to relaxed UMP2-CCSD. Unrelaxed UMP2-CCSD is intermediate in performance between the composite theory and the preferable relaxed UMP2-CCSD approach. 
For relaxed UMP2-CCSD, a significant finding is that we achieve similar accuracy for a fixed active space when the basis set size is increased, which is generally not guaranteed for embedding methods.  

The efficacy of relaxed MP-CCSD has been further established by studying rotational isomerism in 1,3-butadiene. The UMP2 method predicts an anomalous trend in the relative energies of the rotamers, which the unrelaxed version of our method cannot rectify. However, the relaxed UMP-CCSD method can recover the correct trend even with a very small active space.  

We have investigated the reliability of the proposed orbital relaxed UMP-CCSD method for single reference systems in the W4-11 thermochemistry dataset. Our investigations show that relaxed UMP-CCSD reproduces, on average, the UCCSD results to chemical accuracy.

Furthermore, we have studied the central nitrogen double bond stretching of azomethane as a test case of a localized fragment. The relaxed UMP-CCSD method shows an almost constant energy difference relative to the all-electron UCCSD energy along the PEC, whereas the UMP2 method and the unrelaxed UMP-CCSD methods show non-smooth behavior near the symmetry-breaking region of this molecule. 

There is ample scope for future work. The implementation reported and used here is not yet fully optimal, because of an unfavorable $\mathcal{O}(V^2 O^2 O_\mathrm{F} V_\mathrm{F})$ scaling when the renormalized interaction is built before solving the fragment amplitude equations (Fig.~\ref{fig:embed_scheme}). Approximation schemes to construct the renormalized interaction could achieve lower polynomial scaling and also facilitate an efficient multi-fragment approach. This could be an alternative to existing embedding approaches such as LASSCF~\cite{Hermes_LASSCF_JCTC19,hermes2020variational,pham2018can,Agarawal2024Automatic}, DMET~\cite{Knizia12_DMET,dmet_knizia13, Pham_JCTC2018,Booth_PhysRevX22,nusspickel2023effective}, SEET~\cite{Kananenka15, Iskakov20}, DMFT~\cite{Georges96, Kotliar06, LDAplusDMFT_Anisimov1997} or g-RISB~\cite{RISB_PhysRevB19, RISB_PhysRevB23}. 
Moreover, there is also the possibility of utilizing efficient local correlation methods to reduce the complexity further.
With respect to the high-level CC method, it would be very desirable to extend the theory to CCSDT; preliminary analysis has shown that MP-CC has a favorable scaling when including triples amplitudes.
With respect to the low-level MP method, it will be interesting to investigate alternatives to MP2 such as Brillouin-Wigner perturbation theory (BWPT) \cite{BWPTJCP2023}, random phase approximation (RPA) or ring coupled cluster doubles (r-CCD).

\section{Acknowledgements}

This material is based upon work supported by the U.S. Department of Energy, Office of Science, Office of Advanced Scientific Computing Research and Office of Basic Energy Sciences, Scientific Discovery through Advanced Computing (SciDAC) program under Award Number DE‐SC0022198 (A.S., K.B.W. and L.L.), and Award Number DE-SC0022364 (M.H-G.) and by the Simons Targeted Grants in Mathematics and Physical Sciences on Moir\'e Materials Magic (F.F.). L.L. is a Simons Investigator in Mathematics. This research used resources of the National Energy Research Scientific Computing Center, a DOE Office of Science User Facility supported by the Office of Science of the U.S. Department of Energy under Contract No. DE-AC02- 05CH11231 using NERSC award BES-ERCAP0029462.
 A.S. and L.L. thank Karol Kowalski for insightful discussions.
Additionally, the authors thank the anonymous reviewers for their comments, which have contributed to the improvement of this manuscript.

\bibliographystyle{apsrev4-1}
\bibliography{coupled_cluster, embedding, misc}

\pagebreak
\widetext
\appendix

\section{Comparison with the MLCC method}
\label{Sec:complexity_triples}

In the multi-level CC (MLCC) method, Koch and co-workers defined a wave function ansatz 

\begin{equation}
    |\Psi \rangle = e^{X} |\Phi_0\rangle ; \quad X = T_1 + T_2^{act} + T_2^{inact}
\end{equation}
where, $T_1$ is both an active and inactive set of singles amplitudes, $T_2^{act}$, and $T_2^{inact}$ stand for active, and inactive doubles amplitudes. Then to derive the inactive projection equations as in Eq.~\eqref{Eq:projEB}, they assign a perturbative order for $T_2^{inact}$, but the rest are treated in all orders. Furthermore, the projection equation for inactive $T_1$ amplitudes is the same as CC amplitude equations, but the one for $T_2^{inact}$ is truncated at the first order. The projection equations for $T_2$ active amplitudes do not differ from CC. This scheme is similar to using CC2 \cite{CC2CPL1995} equations for the inactive amplitudes. Then, to include triples excitation, they augment X with active triples amplitudes, $T_3^{act}$. Therefore, their treatment of the doubles and triples amplitudes is not on the same footing. With this ansatz their low-level projection equations contain singles and doubles equations, but no triples equations. When triples are included in that work, both for the singles and doubles equations, there is a $[V, T_3]$ term. But because of the active restriction in $T_3$, it maximally scales as $n_{v, act}^3 n_{o, act}^3 n_v$.  

Our projection equations for the active amplitudes are full CC amplitude equations similar to the MLCC method, but the inactive projection equations are derived differently than that method. We do not define a wave function ansatz for the total problem but tried to pick a subset of terms/diagrams from the full set of terms of the CC amplitude equation. This subset of terms is chosen based on a certain heuristic that ensures the low-scaling nature of the inactive amplitude equation. It allows us to choose a projection equation like Eqs.~\eqref{Eq:low_relax_singles} and ~\eqref{Eq:low_relax_doubles}, where we have truncated both the singles and doubles equations in the same order of perturbation, unlike the CC2 projection equations. This choice produces a minimal model that can induce orbital relaxation for the environment when the active space/fragment $T_1$ amplitudes become nonzero. When we include active $T_3$ equations for the fragment, it does not give rise to any new sets of inactive $T_3$ amplitudes because of the lack of $[V, T_3]$ or $[V, T_2]$ terms which are second-order terms (note that, the perturbative order argument here is different than the M{\o}ller-Plesset perturbation series as we treat $T_1$ amplitudes to all orders for the low-level method). Therefore, with our approach when triples are included, the complexity of the low-level equations does not increase compared to the singles-doubles equations.

\section{Working equations for fragment amplitude equation:} \label{Sec:WFEquations}
We have summarized the working equations for the fragment amplitude equation in this section. We will use uppercase letters (I, J, K, L,... for hole type; A, B, C, D,... for particle type)  to denote orbitals in the fragment, and lowercase letters (i, j, k, l,... for hole type; a, b, c, d,... for particle type) for orbitals in the environment. The coupled cluster residue has been denoted as $R$, $f$, $V$ tensors stand for bare interaction, and $F$, $W$ stand for similarity transformed one- and two-particle interactions. We have not shown all the explicit expressions for $F$ and $W$, as they can be found in many coupled cluster literature \cite{stanton_gaussJCP91}.  

\subsection{One-body residue:}

\begin{align}
    R^A_I = {} & F^A_I +  F^A_E t^E_I - F^M_I t^A_M - V^{AM}_{EI} t^E_M + F^M_E t^{AE}_{IM} + \frac{1}{2} V^{AM}_{EF} \tau^{EF}_{IM} - \frac{1}{2} W^{MN}_{EI} t^{EA}_{MN} 
\end{align}
where, 
\begin{align}
     F^A_I = {} & f^A_I + F^A_e t^e_I - F^m_I t^A_m - V^{Am}_{eI} t^e_m + F^m_e t^{Ae}_{Im} +  \frac{1}{2} V^{Am}_{ef} \tau^{ef}_{Im} - \frac{1}{2} W^{mn}_{eI} t^{eA}_{mn} \\
     \tau^{ab}_{ij} = {} & t^{ab}_{ij} + P(ij) \frac{1}{2} t^a_i t^b_j \\
      P(pq) f(p,q) = {} & f(p,q) - f(q,p) 
\end{align}

\subsection{Two-body residue:}
\begin{align}
    R^{AB}_{IJ} = {} & W^{AB}_{IJ} + P(IJ) V^{AB}_{EJ} T^E_I - P(AB) W^{AM}_{IJ} T^B_M + P(AB) F^A_E T^{EB}_{IJ} - P(IJ) F^M_I T^{AB}_{MJ} + \frac{1}{2} V^{AB}_{EF} \tau^{EF}_{IJ} \nonumber \\ 
    {} & + \frac{1}{2} W^{MN}_{IJ} \tau^{AB}_{MN} + P(AB) P(IJ) W^{AM}_{EI} T^{EB}_{JM}   
\end{align}
where,
\begin{align}
     W^{AB}_{IJ} = {} & V^{AB}_{IJ} + P(IJ) V^{AB}_{eJ} T^e_I - P(AB) W^{Am}_{IJ} T^B_m + F^A_e T^{eB}_{IJ} - P(IJ) F^m_I T^{AB}_{mJ} + \frac{1}{2} V^{AB}_{ef} \tau^{ef}_{IJ} \nonumber \\
     {} &  + \frac{1}{2} W^{mn}_{IJ} \tau^{AB}_{mn} + P(AB) P(IJ) W^{Am}_{eI} T^{eB}_{Jm}
\end{align}

The most expensive term in the above amplitude equations is the following one:
\begin{equation}
    W^{Am}_{eI} = 0.5*V^{mn}_{ef}T^{Af}_{nI}
\end{equation}

\section{The choice of basis for the active space:} \label{sec:cmovslmo}
\begin{figure}[ht]
    \centering
    \includegraphics[width=0.45\textwidth]{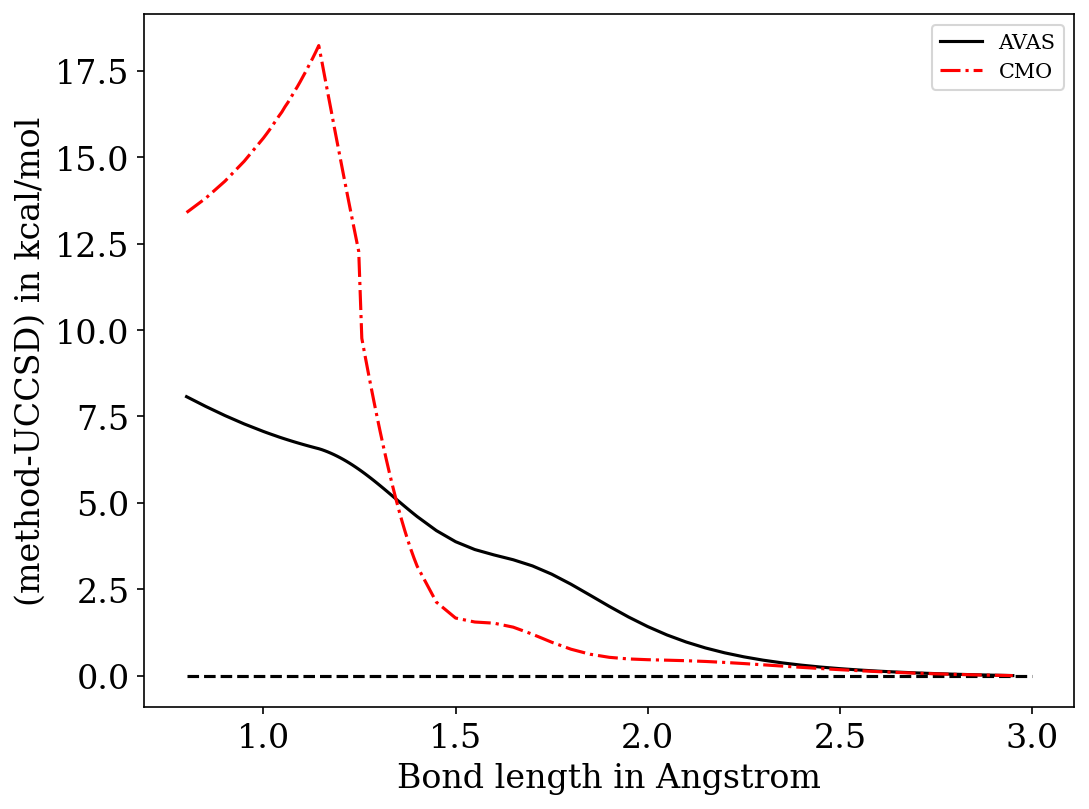}
    \caption{Energy difference between the UCCSD and relaxed UMP-CCSD methods for N$_2$ molecule in aug-cc-pCVDZ basis set. Active space B as described in Table ~\ref{tab:active_space_n2} was considered for this study.}
    \label{fig:active_space_basis}
\end{figure}

We have numerically compared two different choices for the active space basis: canonical molecular orbital (CMO)s, and intrinsic atomic orbital (IAO) basis within the active valence active space (AVAS) scheme. AVAS scheme allows us to choose the active orbitals for a specific fragment of the system. For this comparison, we have chosen the example of N$_2$ molecule as described in \ref{sec:results}. We have used aug-cc-pCVDZ basis set and B-type active space as described in Table ~\ref{tab:active_space_n2} for this study. Our findings are summarized in Fig.~\ref{fig:active_space_basis} where we have employed the relaxed version of our embedding scheme. We found that only the active space chosen with the AVAS scheme can remove the kink near the Coulson-Fischer point.    

\section{Additional plots}

\begin{figure}[ht]
    \centering
    \includegraphics[width=0.45\textwidth]{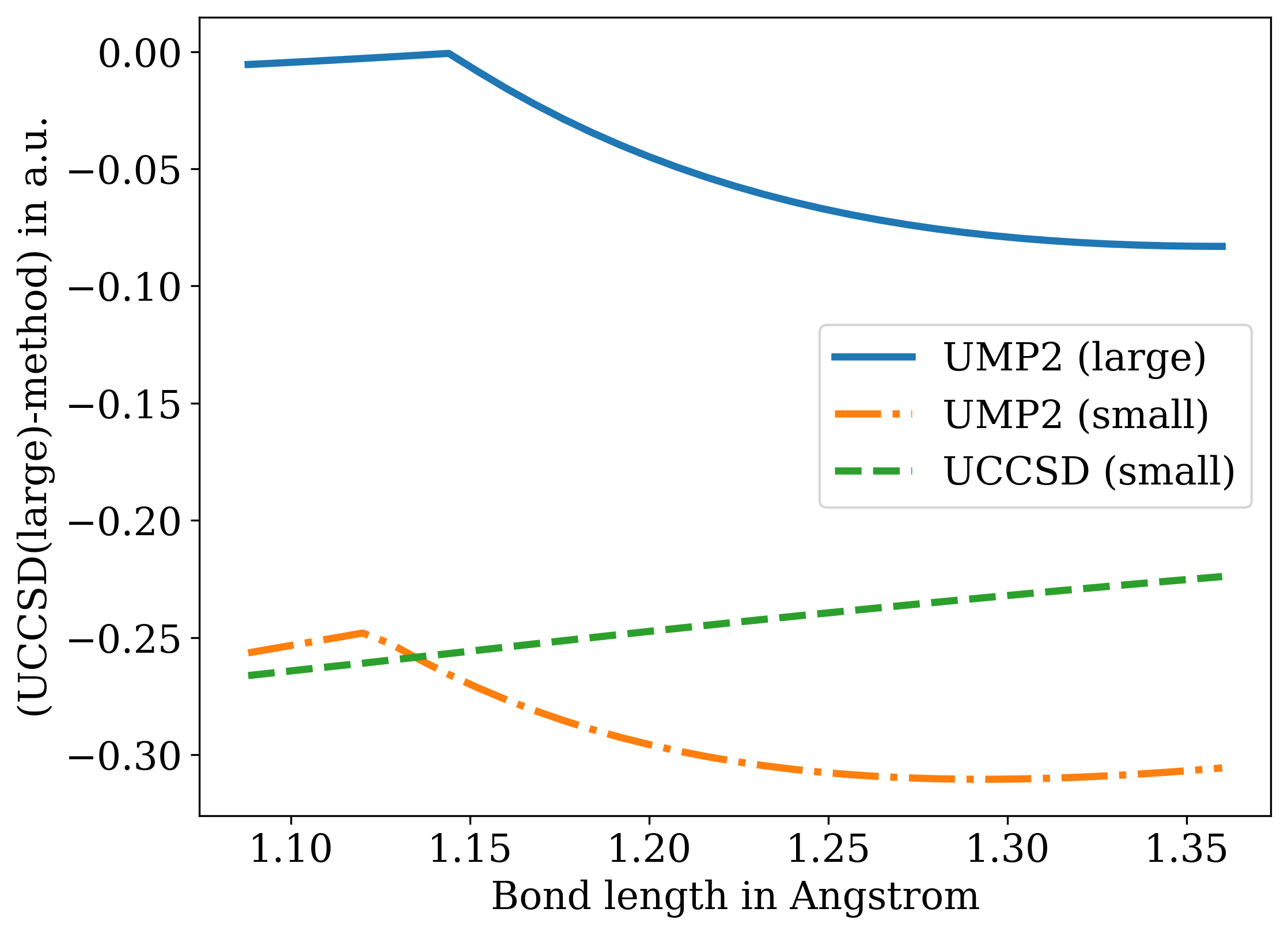}
    \caption{Correlation energy differences w.r.t.~UCCSD in aug-cc-pCVDZ basis set along a PEC for N$_2$. large and small basis sets stand for aug-cc-pCVDZ and SVP respectively.}
    \label{fig:corr_energy_diff_composite_n2}
\end{figure}

\begin{figure}[ht]
    \centering
    \includegraphics[width=0.45\textwidth]{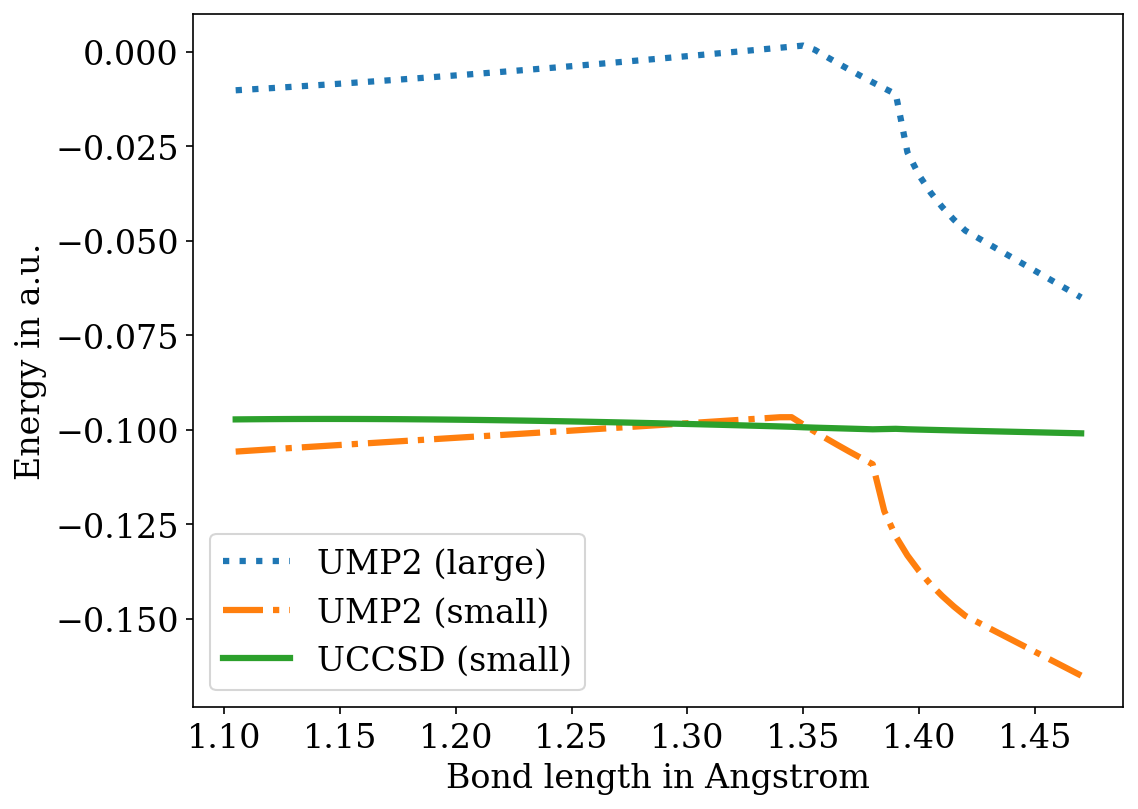}
    \caption{Correlation energy differences w.r.t.~UCCSD in aug-cc-pCVDZ basis set along a PEC for CO. large and small basis sets stand for aug-cc-pCVDZ and cc-pVDZ respectively.}
    \label{fig:corr_energy_diff_composite_co}
\end{figure}

\begin{figure}[ht]
    \centering
    \includegraphics[width=0.5\textwidth]{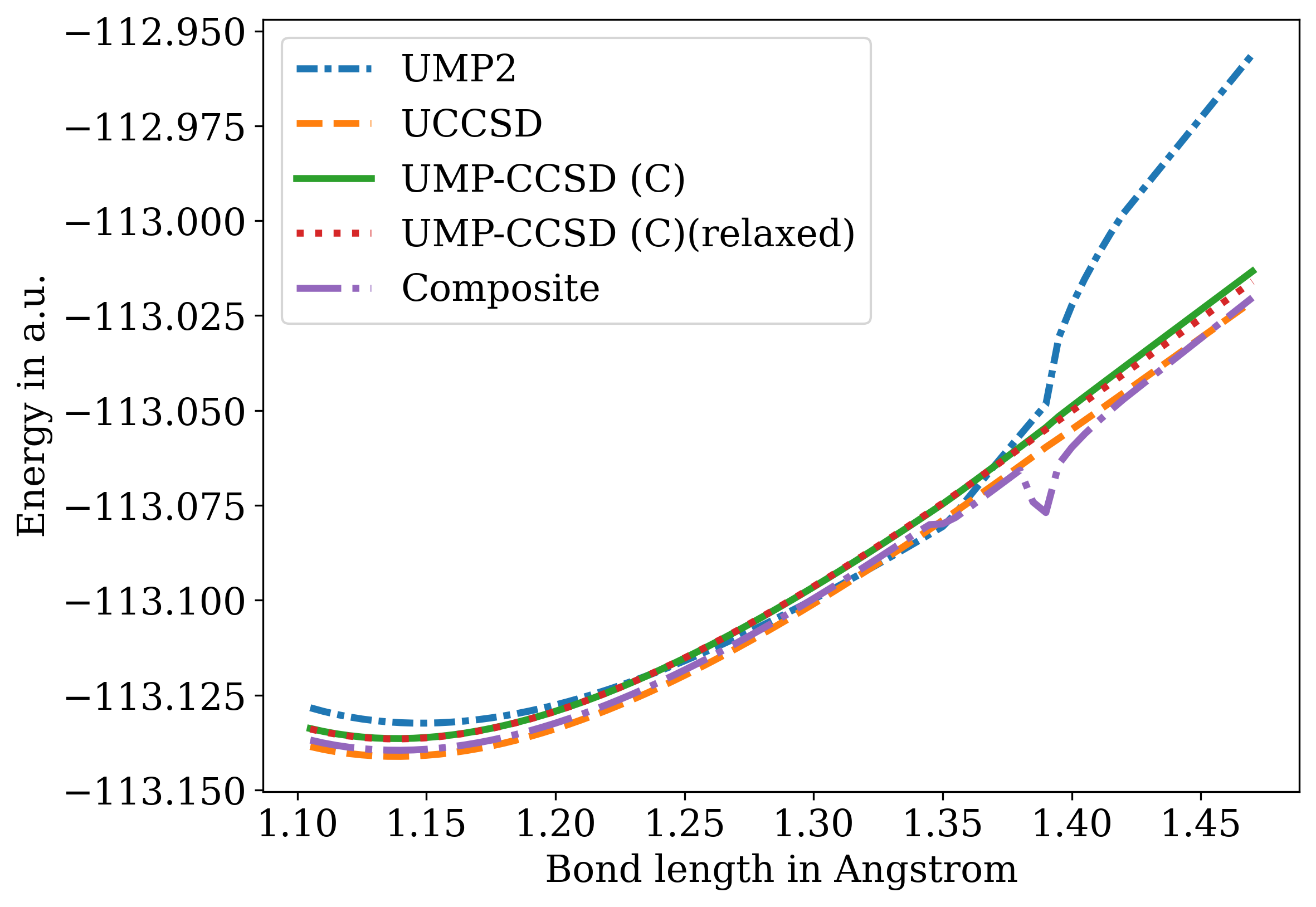}
    \caption{Total energy differences w.r.t.~UCCSD in aug-cc-pCVDZ basis set along a PEC for CO.}
    \label{fig:pec_co}
\end{figure}

We have added Fig.~\ref{fig:corr_energy_diff_composite_n2} and \ref{fig:corr_energy_diff_composite_co} to aid the analysis of composite method over the PEC of $N_2$ and CO respectively. Additionally, Fig.~\ref{fig:pec_co} was added to show the section of PEC of CO where multiple discontinuities arise for the UMP2 method and the composite method.

\newpage

\section{Measure of spin symmetry breaking in 1,3-butadiene}
In this section, we have tabulated the $\langle S^2 \rangle$ value as a measure of spin symmetry breaking in various rotational isomers of 1,3-butadiene molecule. 
\begin{table}[ht]
    \centering
    \begin{tabular}{|c|c|c|}
        \hline
        Isomer & $\langle S^2 \rangle_{UHF}$ & $\langle S^2 \rangle_{UCCSD}$ \\
        \hline
         $\tau = 0^{\circ}$ & 0.450596 & 0.011599 \\
         \hline
         $\tau = 35.5^{\circ}$ & 0.310997 & 0.004780 \\
         \hline
         $\tau = 90^{\circ}$  & 0.003347 & 0.000019 \\
         \hline
         $\tau = 101.7^{\circ}$ & 0.039688 & 0.000247 \\
         \hline
         $\tau = 180^{\circ}$ & 0.384190 & 0.008753 \\
         \hline
    \end{tabular}
    \caption{$\langle S^2 \rangle$ for different rotamers of 1,3-butadiene from UHF and UCCSD methods.}
    \label{tab:s2_butadiene}
\end{table}

\end{document}